\documentclass[12pt]{article}
\pdfoutput=1             
\usepackage{amsmath,amssymb}
\usepackage{graphicx}
\usepackage{jheppub}
\usepackage{tikz}
\usepackage{ytableau}
\usepackage{genyoungtabtikz}
\usetikzlibrary{arrows,shapes,positioning}
\usetikzlibrary{decorations.markings}
\usetikzlibrary{arrows,snakes,backgrounds}
\usepackage{rotating} 
\usetikzlibrary{decorations.markings}
\title{A Young diagram expansion of the hexagonal Wilson loop (amplitude) in ${\cal N}=4$ SYM}

\author[a]{Davide Fioravanti,}
\author[b]{Hasmik Poghosyan,}
\author[b]{Rubik Poghossian}

\affiliation[a]{Sezione INFN di Bologna and
	Dipartimento di Fisica e Astronomia\\ Universit\`a di Bologna,
	Via Irnerio 46, 40126 Bologna, Italy}
\affiliation[b]{Yerevan Physics Institute\\
	Alikhanian Br. 2, 0036 Yerevan, Armenia}
\emailAdd{fioravanti AT bo.infn.it}
\emailAdd{hasmikpoghos@gmail.com}
\emailAdd{poghos@yerphi.am}

\abstract{We shall interpret the null hexagonal Wilson loop (or, equivalently, six gluon scattering amplitude) in 4D ${\cal N}=4$ Super Yang-Mills, or, precisely, an integral representation of its matrix part, via an ADHM-like instanton construction. In this way, we can apply localisation techniques to obtain combinatorial  expressions in terms of Young diagrams. Then, we use our general formula to obtain explicit expressions in several explicit cases. In particular, we discuss those already  available in the literature and find exact agreement. Moreover, we are capable to determine explicitly the denominator (poles) of the matrix part, and find some interesting recursion properties for the residues, as well.}

\begin{document}
\maketitle
\newcommand{\ie}{{\it i.e.\ }}
\def\bea{\begin{eqnarray}}
\def\eea{\end{eqnarray}}

\tikzset{middlearrow/.style={
		decoration={markings,
			mark= at position 0.5 with {\arrow{#1}} ,
		},
		postaction={decorate}
	}
}


\section{Elements of Wilson loops/amplitudes in ${\cal N}=4$ SYM}
Among the different supersymmetric gauge theories a special r\^ole is played by the maximal one in 4D, ${\cal N}=4$ Super Yang-Mills (SYM), with gauge group $SU(N_c)$ and dimensionless coupling constant $g_{YM}$. Needless to say that we are interested in gauge theories for phenomenological reasons and in this particular one because it is conformal, albeit there are at least two other connected motivations. One is that this theory forms one side of the most known example of AdS/CFT correspondence \cite{MGKPW1,MGKPW2,MGKPW3}, as it lives on  the $4d$ Minkowski boundary of its gravitational {\it dual}, the type IIB superstring theory on $AdS_5 \times S^5$. The second reason is that the latter appears to be classically integrable in the sense that it can be written in Lax form \cite{BPR}, and moreover quantum integrable at least in the planar limit, $N_c\rightarrow \infty$ with $g_{YM}\to 0$ so that the 't Hooft coupling $\lambda\equiv N_c g_{YM}^2=16 \pi ^2 g^2$ stays fixed, in that $\mathcal{N}=4$ SYM shows remarkable connections with $1+1$ dimensional integrable models \cite{MZ, BS1,BS2,BS3,BS4,BS5,BES,TBA1,TBA2,TBA3,QSC1,QSC2}. The presence of integrability has opened the way to a better comprehension and partial proof of the aforementioned correspondence.

More precisely, quantum integrability was discovered for the spectral problem of $\mathcal{N}=4$, \emph{i.e.} the computation of the anomalous dimensions of local (single trace) gauge invariant operators. But, more recently, it played an important r\^ole also in the evaluation of null polygonal Wilson loops (Wls), which are dual to and the same as gluon scattering amplitudes \cite{AM-amp, DKS, BHT}; but in a different, though connected, guise. 

In a nutshell, $\mathcal{N}=4$ SYM, as any conformal quantum field theory, enjoys an Operator Product Expansion (OPE) of the (renormalised) null polygonal Wilson loops \cite{Anope}, the simplest ones of non-local nature. This method has an intrinsic non-perturbative nature, given by the collinearity expansion of the null edges which gives rise to an infinite series over the number of particles ({\it cf. infra}). This OPE series was developed for the Wls in $\mathcal{N}=4$ SYM by employing the underlying integrability of the theory, which manifests itself in the flux-tube, dual to the Gubser, Klebanov and Polyakov (GKP) \cite{GKP} string. The connection with the spin chain picture was initiated in \cite{Bel-Ope,Bel-Qua}, but the OPE series for the Wilson loop was developed in a series of papers \cite{BSV1,BSV2,BSV3}, through the so-called \emph{pentagon approach}. Briefly, the proposal was to write the expectation value of the Wl as an infinite sum over intermediate excitations on the GKP string vacuum: gluons and their bound states, fermions, antifermions and, finally, scalars. In this way, it is reminiscent of the Form Factor (FF) spectral (large distance) expansion of the correlation functions in integrable $2d$ quantum field theories, and the pentagonal operator has been identified \cite{BSV1, Bel-Tw1, Bel-Tw2,  CDF} with a specific {\it conical twist} field \cite{Knizhnik:1987xp, CCAD}. Therefore, this methodology needs to know the dispersion relations of the GKP string excitations \cite{Basso} and the $2d$ scattering factors between them  \cite{FRO6, Basso-Rej, FPR1}: these quantities can be gained by filling up (with infinite Bethe roots) the original BPS vacuum of the single trace operators (as ferromagnetic vacuum) to construct the GKP (or anti-ferromagnetic) vacuum and the  excitations thereof. 

The validity of the OPE series has been checked in the weak coupling ({\it e.g.} \cite{BSV2,P1,P2,DP}) and the strong coupling regimes ({\it e.g.} \cite{BSV3,BSV4,FPR2,BFPR2, BFPR1, Bel1509,ISS}), with comparisons against gauge and string theory, respectively. In the latter regime, string theory has so far given only the leading order as minimization of the world-sheet area living in the $AdS_5$ space and insisting on the polygon in $4d$ Minkowski \cite{Anope,TBuA,YSA, Hatsuda:2010cc}. This contribution is given by gluons and (bound states of) fermions in the OPE series \cite{FPR2,BFPR1,proc}. Although scalars elude the minimal area argument, nevertheless they are even more important to understand as they provide a comparable non-perturbative contribution, heuristically derivable from the $O(6)$ non-linear sigma model and corrections as emerging from the low energy action of the string on $S^5$: for the hexagon this has been proven in the FF set-up by \cite{BSV4,Belsca,BFPR2, BFPR3, BFPR4}. In general, their OPE contribution to the $N$-sided polygon is a form factors series of a $(N-4)$-point correlation function of a string model, which reduces to the $O(6)$ non-linear sigma-model in the strong coupling \cite{BSV4, BFPR2, BFPR3, BFPR4}. In this regime the dynamically generated mass is exponentially suppressed $m\sim e^{-\frac{\sqrt{\lambda}}{4}}$ and entails the short-distance regime for the correlation function and thus a power-law behavior \cite{BSV4}. Thanks to the specific form of this contribution at all couplings, it has been possible to expand it with exact computations in this regime by using integrability (only, so far) derived ideas and methods, so proving it to be of the same order as the one coming from the classical string computation\cite{BFPR2, BFPR3, BFPR4}. In this respect, the all coupling form of the contribution of scalars assumes a particular relevance and it will be given a new interpretation in this paper.

In general, if $\mathfrak{u}_i$ are the rapidities of the GKP string (flux tube), the hexagonal Wilson loops can be represented as the OPE  
\bea
\label{Wilson}
W=  \sum_{\mathfrak{n}} \frac{1}{S_{\mathfrak{n}}}\int\prod_{i=1}^{\mathfrak{n}} \frac{d \mathfrak{u}_i}{2\pi}\Pi_{dyn}(\{\mathfrak{u}_i\})\Pi_{FF}(\{\mathfrak{u}_i\})\Pi_{mat}(\{\mathfrak{u}_i\}) .
\eea
over all admissible $\mathfrak{n}$ particle states. The explicit form of $\Pi_{dyn}$ and $\Pi_{FF}$ can be found e.g. in   \cite{Basso:2015rta} and $S_{\mathfrak{n}}$ is the symmetry factor:
given $\mathfrak{n}=\mathfrak{n}_1+\mathfrak{n}_2+...+\mathfrak{n}_k$, where  $\mathfrak{n}_j$ are the number of identical particles of a given kind $j$,  then $S_\mathfrak{n}=\mathfrak{n}_1!\mathfrak{n}_2!...\mathfrak{n}_k!$. In this paper, we will be interested only in $\Pi_{mat}$, which admits an integral representation as multi-integrals over the nested Bethe (or isotopic) roots \cite{FPR1, FPR2} (see below (\ref{gen_Pmat1}) and notice that it depends only  on the rapidities of scalars, fermions and antifermions $\{u_i\}$, $\{v_i\}$, $\{\bar{v}_i\}$). For large number of particles the evaluation of these integrals soon becomes formidable. In fact, our aim here is to represent this integral as a combinatorial sum, which instead allows for explicitly calculation also for a large number of particles.

In papers \cite{BFPR3, BFPR4} one of the authors (D.F.) and collaborators have focused on the matrix part with only scalars, {\it cf.} (\ref{Pmat1}), or only fermions respectivey, {\it cf.} (\ref{fPmat1}), namely as multi-integrals over the nested Bethe roots of  $SO(6)$ or $SU(4)$, respectively. They systematically evaluated by residues the result and encoded it in some Young diagram combinatorics \cite{BFPR2,BFPR3,BFPR4}. This method gives rather explicit final formul\ae \, in terms of simple rational functions and is reminiscent of the pole contributions to the instanton partition function of $\mathcal{N}=2$ SYM \cite{Nekrasov:2002qd}. But still the diagrams are rather different and comparison is not so strict, neither physically motivated. This is why in the present paper, we want to draw a more refined correspondence between the two methods by making explicit and adequate reference to instantons, namely the equivariant localization technique and its related Atiyah-Drinfeld-Hitchin-Manin (ADHM) construction \cite{Atiyah:1978ri}. Importantly, we can in this way treat the most general situation with any number of scalars {\it and} fermions and reach the simple form of sum over multiple Young diagrams of certain shapes, where each term is the inverse of a nice factorized polynomial expression.

In this perspective, the paper is organized as follows: in section \ref{scalars}, we consider the matrix part of MHV amplitudes where one has only $2n$  scalars. Upon taking into account the  structure of the integral representation we construct an appropriate  complex manifold defined by a  system of matrix equations. Then we apply equivariant localization to get a combinatorial representation of the  matrix part of  Wl. Several applications and consequences of our formula are given as well.
In section \ref{fermion}, we extend the above analysis to the case with an equal number of fermions and antifermions, without scalars. Then, in section \ref{Gen_formula},  we consider the matrix part of the Wl with $N_{\phi}$ scalars, $N_{\psi}$ fermions and $N_{\bar{\psi}}$ antifermions. In this general case too, we suggest for the integral representation (\ref{gen_Pmat1}) the matrix equations (\ref{ADHMv})-(\ref{ADHMu}), which, via localization, lead to the combinatorial formula presented in subsection \ref{gen_main}. Eventually, in section \ref{GEN_resulats} we use our general formula (\ref{PIYoung_gen}) to establish the result  (\ref{den}) for the denominator  of $\Pi^{(N_{\psi}N_{\phi}N_{\bar{\psi}})}_{mat}$. We also have found two interesting residue formulae (\ref{rec_gen_sc}) and  (\ref{recF_gen}). At the end of this section for several specific values of scalars, fermion, antifermions and the bottom $R$-charge $r_b$ we derive the matrix parts explicitly.

\section{Scalars}
\label{scalars}
 In this section  we rewrite the integral representation of the  matrix part of the hexagonal bosonic Wl in such a way to make the connection with the integral representation of instanton partition function in  $\Omega$ background more obvious. The structure of integral representation suggests how to construct an ADHM-like moduli space which reproduces the same integral  via equivariant localization. As usual the denominator encodes the required set of algebraic data (matrices) and the numerator reflects the symmetry  properties of equations to be satisfied by these data. Thus,   we introduce the appropriate ADHM-like moduli space defined in terms of six matrices subject to three quadratic equations. It is shown that this moduli space admits a $U(1)^{2n+1}$ symmetry which is used to apply equivariant localization technique. Subsection \ref{possDIAG} is devoted to the classification of fixed points of moduli space under aforementioned $U(1)^{2n+1}$ action. It is shown that each fixed point can be described by an array of $2n$ Young diagrams.  Only six types of diagrams (an empty one, a 1-box, two 2-box, a 3-box and  a 4-box diagrams shown in Fig.\ref{posibleYD}) are allowed. We investigate the tangent space of an arbitrary fixed point and find a closed character formula which encodes the pattern of how a tangent space decomposes into one (complex) dimensional invariant subspaces of the $U(1)^{2n+1}$ action. Finally, we apply localization technique to present the matrix part as a sum over fixed points, {\it i.e.} $2 n$-tuples of above young diagrams constructed in  subsection \ref{possDIAG}. We use our formula to derive explicitly the first $2$ and $4$ particles expressions, $\Pi_{mat}^{(2)}$ and $\Pi_{mat}^{(4)}$ respectively and find complete agreement with all previous results available in literature.

\subsection{Connexion to the ${\cal N}=2$ ($\Omega$ background) SYM partition function}
\label{N2vsN4sym}
As for the hexagonal bosonic Wl in ${\cal N}= 4$ SYM, the matrix factor accounting for the 2D scattering of $2n$ scalars (of GKP string or flux tube) does not depend on the 't Hooft coupling constant $\lambda$ and can be written as a multi-integral over the three kinds of nested Bethe Ansatz roots of a $SO(6)$ spin-chain \cite{FPR1, FPR2}, $a_k$, $b_j$ and $c_k$ \cite{BSV4,BFPR2}:
\bea
\label{Pmat1}
\Pi_{mat}^{(2n)}(u_1,....,u_{2n})&=&\frac{1}{(2n)!(n!)^2}
\int_{-\infty}^{\infty}
\prod_{k=1}^{n}\frac{d\,a_k}{2\pi}\prod_{j=1}^{2n}\frac{d\,b_j}{2\pi}
\prod_{k=1}^{n}\frac{d\,c_k}{2\pi}\times\\ \nonumber
&\times&\frac{\prod^{n}_{i<j}g(a_{ij})\prod^{2n}_{i<j}g(b_{ij})\prod^{n}_{i<j}g(c_{ij})}
{\prod_{j=1}^{2n}\left(\prod_{i=1}^{n}f(a_i-b_j)\prod_{k=1}^{n}f(c_k-b_j)\prod_{l=1}^{2n}f(u_l-b_j)\right)}\,,
\eea
where the functions $f(x)$ and $g(x)$ are defined as
\bea
\label{f_g_def}
f(x)=x^2+1/4=(x-i/2)(x+i/2)\,,
\,\,
g(x)=x^2(x^2+1)=x^2(x-i)(x+i)\,.
\eea
Inserting this in (\ref{Pmat1}) we get 
\bea 
&\Pi_{mat}^{(2n)}(u_1,....,u_{2n})=\frac{1}{(2n)!(n!)^2}
\int_{-\infty}^{\infty}
\prod_{k=1}^{n}\frac{d\,a_k d\,c_k}{(2\pi i)^2}\prod_{k=1}^{2n}\frac{d\,b_k}{2\pi i}
\times \qquad \qquad\qquad\qquad\\ \nonumber
&\qquad\times\frac{\prod^{'n}_{	i,j=1
	}a_{ij}c_{ij}
	(a_{ij}+i)(c_{ij}+i)	
	\prod^{'2n}_{ 	\alpha ,\beta=1
	}b_{\alpha\beta}(b_{\alpha\beta}+i)}{\prod_{\alpha=1}^{2n}\prod_{i=1}^{n}
(a_i-b_{\alpha}+{i\over2})(b_{\alpha}-a_i+{i\over2})(c_i-b_{\alpha}+{i\over2})(b_{\alpha}-c_i+{i\over2})
\prod_{\ell=1}^{2n}\prod_{\alpha=1}^{2n}(u_{\ell}-b_{\alpha}+{i\over2})(b_{\alpha}-u_{\ell}+{i\over2})}\,,
 \eea 
 where the prime on the product symbol indicates that all the vanishing factors ({\it i.e.} the factors $a_{ii}$, $c_{ii}$,
 $b_{\alpha,\alpha}$) are omitted. After generalizing the last expression by the substitution $\frac{i}{2}\rightarrow \epsilon$ and shifting the variables of integration by this amount we get
\bea \label{Pmat2}
&\Pi_{mat}^{(2n)}(u_1,....,u_{2n})=\frac{1}{(2n)!(n!)^2}
\int_{-\infty}^{\infty}
\prod_{k=1}^{n}\frac{d\,a_k d\,c_k}{(2\pi i)^2}\prod_{k=1}^{2n}\frac{d\,b_k}{2\pi i}
\times\qquad \qquad\qquad\qquad\\ \nonumber
&\qquad\times \frac{\prod^{'n}_{ 	i,j=1
	}a_{ij}c_{ij}
	(a_{ij}+2\epsilon)(c_{ij}+2\epsilon)\prod^{'2n}_{	\alpha ,\beta=1
	}b_{\alpha\beta}(b_{\alpha\beta}+2\epsilon)}{\prod_{\alpha=1}^{2n}\prod_{i=1}^{n}
	(a_i-b_{\alpha}+\epsilon)(b_{\alpha}-a_i+\epsilon)(c_i-b_{\alpha}+\epsilon)(b_{\alpha}-c_i+\epsilon)
	\prod_{\ell=1}^{2n}\prod_{\alpha=1}^{2n}(u_{\ell}-b_{\alpha})(b_{\alpha}-u_{\ell}+2\epsilon)}\,.
\eea 
A    similar integral arises in a completely different context, the $N = 2$ SYM theory with gauge group $U(n)$ in $\Omega$-background
parameterized by $\epsilon_1$ and $\epsilon_2$. In the latter case, the $k$-instanton contribution to the partition function can be written in the form \cite{Lossev:1997bz,Nekrasov:2002qd} 
\begin{small}
	 \bea \label{inspart}
	Z_k=\frac{1}{k!}
	\oint
	\prod_{l=1}^{k}\frac{d\,\phi_l}{2\pi i}
	\frac{	\prod^{'k}_{	i,j=1
		}\phi_{ij}(\phi_{ij}+\epsilon_1+\epsilon_2)}
	{\prod^{k}_{ 	i,j=1
		}(\phi_{ij}+\epsilon_1)(\phi_{ij}+\epsilon_2)
		\prod_{i=1}^{k}\prod_{\ell=1}^{n}(\phi_i-a_{\ell}+\epsilon_1+\epsilon_2)(\phi_i-a_{\ell})}\,,
	\eea
\end{small}where $a_{\ell}$ are the gauge expectation values and the prime on the product symbol again  means 
  that the  vanishing factors should be suppressed. 
 In the case  $\epsilon_1=\epsilon_2=\epsilon$ this expression is  suspiciously similar to (\ref{Pmat2}).
 On the other hand, upon applying localization techniques for the moduli space of instantons, according to the ADHM construction, the full instanton partition function 
\begin{eqnarray}
Z_{inst}=\sum_{k} Z_k q^k,
 \label{{inspart-full}}
 \end{eqnarray}
can be arranged as a sum over Young diagrams  \cite{Flume:2002az} in the following way
 \begin{eqnarray}
 \label{nekP}
 Z_{inst}=\sum_{\vec{Y}}f_{\vec{Y}}q^{|\vec{Y}|},
 \end{eqnarray}
 where $\vec{Y}$ is an array of  $n$ Young diagrams $\vec{Y}=\{Y_1,Y_2,\dots,Y_n\}$ and 
 $|\vec{Y}|$ is the total
 number of  boxes.  $q$ is the 
 instanton counting parameter, related to
 the gauge coupling constant in the standard manner: $q=\exp 2\pi i 
 \tau $, with  $\tau=\frac{i}{g^2}+\frac{\theta}{2\pi}$ the complexified coupling. The coefficients $f_{\vec{Y}}$ are factorized 
 as
 \bea
 f_{\vec{Y}}=\prod_{u=1}^n\prod_{v=1}^n
 \frac{1}
 {Z_{bf}(a_u,Y_u \mid  a_v,Y_v)}\,.
 \label{f}
 \eea
  This formula can be  obtained applying the so called localization technique for moduli space of instantons.
 Due to the similarities of the contour integral representation of instanton partition function (\ref{inspart}) and the matrix part of the hexagonal bosonic Wilson loop (\ref{Pmat2}) it is reasonable   to expect that there exists a  combinatorial expression like (\ref{nekP}) also for $\Pi_{mat}$. 
 In the next two subsections we will find an ADHM-like construction  such that the corresponding localization formula (see (\ref{PIYoung}), (\ref{Fwil}) and   (\ref{ZbfWil})) gives results compatible with (\ref{Pmat2}).
 \subsection{Matrix equations for the  hexagonal Wilson loop}
 \label{matrix_eq}
 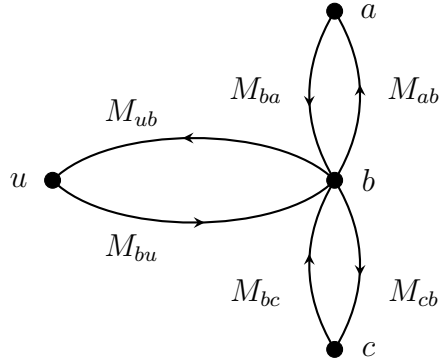
\begin{figure}
	\begin{center}
		\begin{tikzpicture} [scale=1.50]
		\filldraw [black] (0.5,0) circle (2pt)
		(3,0) circle (2pt) 	(3,1.5) circle (2pt) (3,-1.5) circle (2pt);
		\draw[thick, middlearrow={stealth reversed}] (0.5,0) .. controls (1,0.5) and (2.5,0.5) .. (3,0);
		\draw[ thick, middlearrow={stealth reversed}] (3,0) .. controls (2.5,-0.5) and (1,-0.5) .. (0.5,0);
		\draw[ thick, middlearrow={stealth reversed}] (3,0) .. controls (2.7,0.5) and (2.7,1) .. (3,1.5);
		\draw[thick, middlearrow={stealth reversed}] (3,1.5) .. controls (3.3,1) and (3.3,0.5) .. (3,0);
		\draw[thick, middlearrow={stealth reversed}] (3,0) .. controls (2.7,-0.5) and (2.7,-1) .. (3,-1.5);
		\draw[ thick, middlearrow={stealth reversed}] (3,-1.5) .. controls (3.3,-1) and (3.3,-0.5) .. (3,0);
		\node at (0.2,0) {$u$};
		\node at (3.3,0) {$b$};
		\node at (3.3,1.5) {$a$};
		\node at (3.3,-1.5) {$c$};
		\node at (1.2,0.6) {$M_{u b}$};
		\node at (1.2,-0.6) {$M_{ b u}$};
		\node at (2.3,0.8) {$M_{ba}$};
		\node at (3.7,0.8) {$M_{ab}$};
		\node at (2.3,-1) {$M_{bc}$};
		\node at (3.7,-1) {$M_{cb}$};
		\end{tikzpicture}
	\end{center}
	\caption{represents a  quiver diagram where the arrows indicate the linear maps $M_{ab}$, $M_{ba}$, $M_{bc}$, $M_{cb}$,  $M_{bu}$, $M_{u b}$ and dots the spaces $u$, $b$, $a$, $c$ on which they act.}
	\label{BWilsonquiver}
\end{figure}
A nice exposition on how  ADHM  construction leads to the integral representation can be found in \cite{Nekrasov:2004vw}. Here we consider the opposite problem to obtain an ADHM like construction corresponding to (\ref{Pmat2}). 
The ADHM analogy suggests that  we have to introduce four  vector spaces
\begin{itemize}
	\item $u$:  a $2n$ complex dimensional space  related to the parameters $u_1,u_2,\ldots ,u_{2n}$
	\item $b$: a $2n$ complex dimensional space related to the parameters  $b_1,b_2,\ldots ,b_{2n}$
	\item $a$ and $c$: each   a  $n$ complex dimensional space related to the parameters
	$a_1,a_2,\ldots ,a_{n}$ and 	$c_1,c_2,\ldots ,c_{n}$ respectively
\end{itemize}
and six linear  maps $M_{ab}$, $M_{ba}$, $M_{bc}$, $M_{cb}$,  $M_{bu}$, $M_{u b}$ acting  among these spaces as shown in quiver diagram  Fig.\ref{BWilsonquiver}.
The choice of quiver diagram is dictated by the factors in  denominator of Fig.\ref{BWilsonquiver}.  Notice that $M_{bu}$ and $M_{u b}$ are $2n \times 2n$
matrices, $M_{ab}$ and $M_{cb}$ are $n \times 2n$
matrices and finally $M_{ba}$ and $M_{bc}$ are $2n \times n$
matrices. 

The factors $a_{ij}+2\epsilon$, $c_{ij}+2\epsilon$ and  $b_{\alpha \beta}+2\epsilon$ in the  numerator of integral representation (\ref{Pmat2}) suggests that there must be three matrix equations (analogs of complex ADHM), 
having the structure of operators acting as  
 $a\to a$, $c\to c$, $b\to b$ respectively. A simple choice  which eventually  leads to results consistent with the integral representation is
\bea \label{ADHMa}
&M_{ab}M_{ba}=0\;,\\\label{ADHMc}
&M_{cb}M_{bc}=0\;,\\\label{ADHMd}
&M_{ba}M_{ab}-M_{bc}M_{cb}+M_{bu}M_{u b}=0\,.
\eea
As usual in order to guaranty smoothness of moduli space these equations are supplemented by 
  stability condition  \cite{Nakajima:1999vw}. 
  
  We found that the suitable  \textit{Stability condition} reads: There in no proper subspace of $b$ which contains the image $M_{bu}$ and is invariant under two operators  $M_{ba}M_{ab}$ and  $M_{bc}M_{cb}$. In addition it is required that $M_{ab}M_{bu} $ and $M_{cb}M_{bu} $ are \textit{onto} maps on $a$, $c$ respectively.
The first condition means that  if one acts in all possible ways on the image of $M_{bu}$  the entire space $b$ will be recovered. 

Let us introduce also transformations  $T_a$, $T_b$, $T_c$, $T_u$ acting in respective spaces (e.g $T_a:a\to a$ is a $n\times n$ matrix). While $T_u$ is a genuine symmetry (which is the analog of global gauge transformations in $N=2$ SYM), the remaining $T_a$, $T_b$, $T_c$ are auxiliary  transformations (the moduli space is found by factorizing space of solutions over these auxiliary transformations). In addition we introduce also transformation $T_{\epsilon}\in C^*$  ($C^*$ is the group of multiplication by nonzero complex numbers). Let us quote below the transformation lows of matrices $M$    
\bea
\label{tl}
&&M_{ab}\to T_{\epsilon}T_a M_{ab} T_b^{-1}\,, \quad
M_{ba}\to T_{\epsilon} T_b M_{ba} T_a^{-1}\,,\\
&&M_{cb}\to T_{\epsilon} T_c M_{cb} T_b^{-1}\,, \quad
M_{bc}\to T_{\epsilon}T_b M_{bc} T_c^{-1}\,,\\
&&M_{bu}\to T^2_{\epsilon}T_b M_{bu}T_u^{-1}\,, \quad
M_{u b} \to T_u M_{u b} T_b^{-1}\,.
\eea
Notice that the above rules follow the pattern of factors in denominator of (\ref{Pmat2}) (e.g. the first  transformation rule matches with the factor $(a_i-b_{\alpha}+\epsilon)$). \\
\textit{Moduli space}: by definition a point in moduli space ${\cal M}_n$ is a set of matrices \\
$\{ M_{ab},M_{ba},M_{bc},M_{cb},M_{bu},M_{u b}\}$ satisfying the equations (\ref{ADHMa})-(\ref{ADHMd})  and the stability condition together with equivalence relation 
 \bea
 \label{eqrel}
 \{ M_{ab},&M_{ba}&,M_{bc},M_{cb},M_{bu},M_{u b}\} \sim\\ \nonumber
&& \{ T_a M_{ab} T_b^{-1}, T_b M_{ba} T_a^{-1},T_b M_{bc} T_c^{-1}, T_c M_{cb} T_b^{-1},T_b M_{bu}, M_{u b} T_b^{-1}\}\,.
 \eea
It is straightforward to check that the matrix equations (\ref{ADHMa})-(\ref{ADHMd}) respect these transformations. The left hand sides of matrix equations get transform as 
\bea \label{TAADHMa}
&M_{ab}M_{ba}\to T^2_{\epsilon} T_a M_{ab}M_{ba}T_a^{-1},\\\label{TCADHMc}
&M_{cb}M_{bc}\to T^2_{\epsilon} T_c M_{cb}M_{bc} T_c^{-1},\\\label{TBADHMd}
&M_{ba}M_{ab}-M_{bc}M_{cb}+M_{bu }M_{u b}\to T^2_{\epsilon} T_b (M_{ba}M_{ab}-M_{bc}M_{cb}+M_{b u }M_{u b}) T_b^{-1}
\eea
The equations (\ref{ADHMa})-(\ref{ADHMd}) are designed so that their transformation lows exactly much with corresponding factors in the numerators of (\ref{Pmat2}). For example  the equation  (\ref{ADHMa}) transforms as (\ref{TAADHMa}) which obviously agrees with the factor  $a_{ij}+2\epsilon$ in (\ref{Pmat2}).
 
  \subsection{The set of all admissible  diagrams}
  \label{possDIAG}
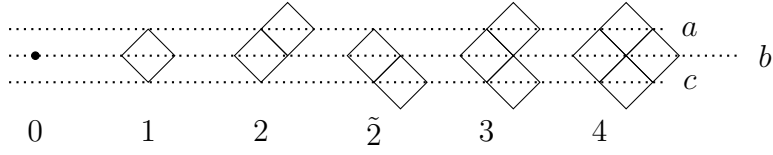
\begin{figure}
	\begin{center}
		\begin{tikzpicture}[scale=0.5]
		\draw [fill] (-1.5,0) circle [radius=0.1];
		\draw (-1.5,-2) node{$0$};
		\draw (1.5,-2) node{$1$};
		\draw (4.5,-2) node{$2$};
		\draw (7.5,-2) node{$\tilde{2}$	};
		\draw (10.5,-2) node{$3$	};
		\draw (13.5,-2) node{$4$	};
	\draw[rotate around={-45:(1.5,0)}] (1,-0.5) rectangle (2,0.5) ;
	\draw [rotate around={-45:(4.5,0)}](4,-0.5) rectangle (5,0.5);
	\draw[rotate around={-45:(4.5,0)}] (4,0.5) rectangle (5,1.5);
	\draw[rotate around={-45:(7.5,0)}] (7,-0.5) rectangle (8,0.5);
	\draw[rotate around={-45:(7.5,0)}] (8,-0.5) rectangle (9,0.5);	
	\draw [rotate around={-45:(10.5,0)}](10,-0.5) rectangle (11,0.5);
	\draw [rotate around={-45:(10.5,0)}]  (10,0.5) rectangle (11,1.5);
	\draw [rotate around={-45:(10.5,0)}] (11,-0.5) rectangle (12,0.5);	
	\draw [rotate around={-45:(13.5,0)}] (13,-0.5) rectangle (14,0.5);
	\draw [rotate around={-45:(13.5,0)}] (13,0.5) rectangle (14,1.5);
	\draw [rotate around={-45:(13.5,0)}] (14,-0.5) rectangle (15,0.5);
	\draw  [rotate around={-45:(13.5,0)}](14,0.5) rectangle (15,1.5);	
	\draw [thick,dotted] (-2.2,0) -- (17.2,0);
	\draw [thick,dotted] (-2.2,0.707106) -- (15.2,0.707106);
	\draw [thick,dotted] (-2.2,-0.707106) -- (15.2,-0.707106);
	\draw (17.9,0) node{$b$};
	\draw (15.9,0.707106) node{$a$};
	\draw (15.9,-0.707106) node{$c$};
		\end{tikzpicture}
	\end{center}
	\caption{The list of allowed diagrams. The diagrams are labeled by $0,1,2,\tilde{2},3,4$ which indicate the number of boxes. Three dotted lines correspond to the spaces $a$, $b$, $c$ such that a box lying  on one of these lines represents a basis vector of the respective space.}
	 \label{posibleYD}
\end{figure}

  To apply localisation we  need to find all fixed points  under transformations $T_{\epsilon}$, $T_u$. Thus a  fixed point is  a set of matrices $\{ M_{ab},M_{ba},M_{bc},M_{cb},M_{bu},M_{u b}\}$ satisfying  the  equations (\ref{ADHMa})-(\ref{ADHMd}) and stability condition, invariant  under  $T_{\epsilon}$, $T_u$   up to auxiliary transformations
  $T_a$, $T_b$, $T_c$.
  It is possible to show that at fixed points the matrix $ M_{u b} =0$,
  so that due to (\ref{ADHMd}) 
  \bea \label{adhmsta}
  M_{ba}M_{ab}=M_{bc}M_{cb}\,.
  \eea
  Let us choose the  basis vectors $e_1, e_2, \dots, e_{2 n}$ in space  $u$ to be eigenvectors of transformation $T_u$:
  \bea
  \label{ei}
  T_ue_i=T_{u_i}e_i\,.
  \eea 
  The we introduce a useful graphical interpretation as follows:   
   \begin{itemize}
   	\item 	
   To each	$e_i$ such that $M_{bu}e_i=0$ 
     we associate a dot (empty  diagram). 
     \item
   	 To each	$e_i$ such that $M_{bu}e_i\ne 0\equiv v$  we associate a box \YRussian \gyoung(v). Thus a box represents a non zero basis vector in space $b$.
   	 \item
   	If   $M_{ab}v_i\ne 0$  then we can add a extra box (NE direction)
   	  \YRussian \gyoung(v;), notice this new box represents a vector in 	the  space $a$.
   	  \item
   	 If   $M_{ba} M_{ab}v_i\ne0$ we can  add a third  box (in SE direction)
   	\YRussian \gyoung(:;,v;), since we have  (\ref{adhmsta}) 
   	 $M_{ba}M_{ab}v=M_{bc}M_{cb}v$  we must add also a box representing $M_{cb}v$
   	 this results  \YRussian \gyoung(;;,v;)
   	 \item If $M_{ab}v_i\neq 0$ and $M_{cb}v_i\neq 0$ then it gives rise to   \YRussian \gyoung(;,v;), 
   	\item Because the spaces  $a$ and $c$ enter our construction in a symmetric way so obviously we have in addition the diagram   \gyoung(;,v).
   \end{itemize}
Due to the matrix equations (\ref{ADHMa}) and (\ref{ADHMc}) it is not possible to enlarge  these diagrams further on.
So we conclude that the diagrams listed in  Fig.\ref{posibleYD} are the only possible ones.

Since the dimension of $u$ is $2n$ we should have  $2n$  diagrams chosen from the list depicted in Fig.\ref{posibleYD}. Let $n_i$ be the number of diagrams of type $i$, $i=0,1,2,\tilde{2},3,4$, then 
\bea 
n_4+n_3+n_2+n_{\tilde{2}} +n_1+n_0=2n\,,
\eea
 The dimensions of $b$, $a$ and $c$ are $2n$, $n$ and $n$ respectively, so 
\bea\label{concba}
2n_4+n_3+n_2+n_{\tilde{2}}+n_1=2n,\,\,\,\,
n_4+n_3+n_2=n,\,\,\,\,
n_4+n_3+n_{\tilde{2}}=n\,.
\eea
 The last four conditions  lead to 
\bea
\label{fincon}
n_4=n_0\,,\quad
n_1=n_3\,,\quad
n_2=n_{\tilde{2}}\,,\quad
n_0+n_1+n_2=n\,.
\eea
\subsection{Tangent space decomposition}
\label{character}
To apply localization formula we have to describe  the tangent spaces of ${\cal M}_n$ at the fixed points.
The procedure is standard:  from the total space where the unconstrained variation $\delta M$ lives, one ``subtracts" subspaces corresponding  to equations (\ref{ADHMa})-(\ref{ADHMd}) and auxiliary transformations. Notice that ({\it cf.} transformation laws (\ref{tl}))
\bea
&\delta M_{ab}\in T_{\epsilon} a \otimes b^{*}\,, \quad
\delta M_{ba}\in T_{\epsilon} b \otimes a^{*}\,\\
&\delta M_{cb}\in T_{\epsilon} c \otimes b^{*}\,, \quad
\delta M_{bc}\in T_{\epsilon} b \otimes c^{*}\,\\
&\delta M_{b u}\in T_{\epsilon}^2 b \otimes u^{*}\,, \quad
\delta M_{u b}\in  u \otimes b^{*}\,.
\eea
Obviously the subspace corresponding to the three  equations (\ref{ADHMa})-(\ref{ADHMd}) is the direct sum of three terms
\bea
\label{eq_sp}
& T_{\epsilon}^2 a \otimes a^{*}\oplus T_{\epsilon}^2c \otimes  c^{*}\oplus  T_{\epsilon}^2 b \otimes b^{*}\,.
\eea
Finally the remaining part corresponding to auxiliary transformations $T_a\in a\otimes a^{*}$, $T_b\in b\otimes b^{*}$, $T_c\in  c\otimes c^{*}$ simply is
\bea
\label{aux_sp}
& a\otimes a^{*}\oplus  b\otimes b^{*}\oplus  c\otimes c^{*}\,.
\eea
 Combining all together for the tangent space character we get
\bea \label{wilchar2}
&\chi_{\vec{Y}} = T_{\epsilon}\left((a+c) b^{*} +b (a^{*}+c^{*})\right)+T_{\epsilon}^2 b u^{*}
+u  b^{*}-\\ \nonumber
&\qquad \qquad\qquad \qquad-(T_{\epsilon}^2+1)\left(a a^{*} + b b^{*}+c c^{*}\right)\,,
\eea
where (without changing notation) we have replaced the spaces by their character (with respect to $T_{\epsilon}$, $T_u$ transformations) as follows:
the character of  space $u$ can be written  as (see (\ref{ei}))
\bea
\label{uch}
u=T_{u_1}+T_{u_2}+T_{u_3}+...+T_{u_{2n}}\,,\quad
u^{*}=T^{-1}_{u_1}+T^{-1}_{u_2}+T^{-1}_{u_3}+...+T^{-1}_{u_{2n}}
\eea
as for the spaces $a$, $b$ and $c$ we will have 
\bea
\label{abcch}
a=\sum^{2n}_{i=1}a_i\,,\quad c=\sum^{2n}_{i=1}c_i\,,\quad b=\sum^{2n}_{i=1}b_i\,,
\eea
where
\bea
&a_k=
\begin{cases}
 T_{\epsilon} T_{u_k }      & \quad \text{if } Y=2,3,4\\
	0  & \quad \text{if } Y=0,1,\tilde{2}
\end{cases}
\,, \quad
c_k=
\begin{cases}
	T_{\epsilon}T_{u_k}      & \quad \text{if } Y=\tilde{2},3,4\\
	0  & \quad \text{if } Y=0,1,2
\end{cases}
\nonumber\\
&b_k=
\begin{cases}
	 T_{u_k}     & \quad \text{if } Y=1,2,\tilde{2},3\\
	(T_{\epsilon}^2+1) T_{u_k }      & \quad \text{if } Y=4\\
	0  & \quad \text{if } Y=0
\end{cases}
\label{a_c_b}
\eea
Notice that since the spaces $a$ and $c$ are $n$ dimensional exactly $n$ therms  in each  (\ref{abcch}) must be zero.
The characters $a^*$, $b^*$, $c^*$ can be found by replacing $a_i \to a_i^{-1}$, $b_i \to b_i^{-1}$, $c_i \to c_i^{-1}$.
For a given fixed point specified by diagrams $\vec{Y}=\{Y_1,Y_2,...,Y_{2n}\}$ the character has the following structure
\bea
\label{X}
\chi_{\vec{Y}}=\sum^{2n}_{i,j=1}X\left(T_{u_i},Y_i|T_{u_j},Y_j\right)\,.
\eea
The summands  $X\left(u_i,Y_i|u_j,Y_j\right)$ can be easily read off from (\ref{wilchar2}):
\begin{small}
	\bea\nonumber
	&X\left(T_{u_i},Y_i|T_{u_j},Y_j\right)=T_{\epsilon}\left(\frac{a_i}{b_j}+\frac{c_i}{b_j}+\frac{b_i}{a_j}+\frac{b_i}{c_j}\right)+T_{\epsilon}^2T^{-1}_{u_j}b_i+\frac{T_{u_i}}{b_j}-\left(T_{\epsilon}^2+1\right)\left(\frac{a_i}{a_j}+\frac{b_i}{b_j}+\frac{c_i}{c_j}\right).
	\quad
	\eea
\end{small}
For convenience all  $X\left(T_{u_i},Y_i|T_{u_j},Y_j\right)$'s are  listed in Table \ref{CHtb}. 
\begin{table}
\begin{center}
	\begin{tabular}{l||cccccc}
		&$0$&$1$&$2$&$\tilde{2}$&$3$&$4$\\
		\hline\hline	
		$0$&	 $0$ & $1$ & $1$ & $1$ & $1$ & $1+T_{\epsilon}^{-2}$  \\
		$1$&	 $T_{\epsilon}^2$ & $0$ & $1$ & $1$ & $2$ & $1$ \\
		$2$&    $T_{\epsilon}^2$ & $T_{\epsilon}^2$ & $0$ & $T_{\epsilon}^2+1$ & $1$ & $1$ \\
		$\tilde{2}$&$T_{\epsilon}^2$ & $T_{\epsilon}^2$ & $T_{\epsilon}^2+1$ & $0$ & $1$ & $1$  \\
		$3$&	$ T_{\epsilon}^2$ & $2 T_{\epsilon}^2$ & $T_{\epsilon}^2 $& $T_{\epsilon}^2$ & $0$ & $1$  \\
		$4$&	 $T_{\epsilon}^4+T_{\epsilon}^2$ & $T_{\epsilon}^2$ & $T_{\epsilon}^2$ & $T_{\epsilon}^2$ & $T_{\epsilon}^2$ & $0$  
	\end{tabular}
\end{center}
\caption{	
	This table gives $X\left(T_{u_i},Y_i|T_{u_j},Y_j\right)$ the factor $T_{u_i}T^{-1}_{u_j}$ is suppressed for example
$X(u_i,4|u_j,4)=0$,
$X(u_i,4|u_j,3)=T_{\epsilon}^2T_{u_i}T^{-1}_{u_j}$ and
$X(u_i,4|u_j,0)=(T_{\epsilon}^2+T_{\epsilon}^4)T_{u_i}T^{-1}_{u_j}$.
}
\label{CHtb}
\end{table}
 It is essential that though  (\ref{wilchar2})  besides positive terms includes also negative ones, nevertheless in final formula (\ref{X}) and Table \ref{CHtb} all terms are positive, which is required by self consistency (the initial space contains all the sub spaces to be factored out due to matrix equations and equivalence relation (\ref{eqrel})). Though we did not provide a mathematically rigorous proof, the above property strongly suggests that the moduli space ${\cal M}_n$ is a {\it smooth} algebraic manifold with dimension $4 n^2$. 
\subsection{Matrix part as a sum over diagrams}
\label{main}
For given $n$ the matrix part of the hexagonal Wl can be represented as  
\bea \label{PIYoung}
\Pi_{mat}^{(2n)}=\sum_{\vec{Y}}F_{\vec{Y}}\,,
\eea
where the  sum is  over all collections of $2n$ diagrams  $\vec{Y}\equiv \{Y_1,Y_2,Y_3,....,Y_{2n}\}$ from the list given in Fig. \ref{posibleYD} subject to constraints: 
\begin{itemize}
	\item The total number of boxes in middle line  (related to the  space $b$) is equal to  $2n$.
	\item Both  the upper and lower lines (related to spaces  $a$ and  $c$) contain    $n$ boxes  each.
\end{itemize} 
A little consideration ensures that above conditions are equivalent   to the requirements  (\ref{fincon}) where  $n_i$ ($i=0,1,2,\tilde{2},3,4$) is the number of diagrams of type $i$ (see Fig.\ref{posibleYD}) in $\vec{Y}$. We can also deduce that if a diagram of type $0$ ($1$ and $2$) enters in $\vec{Y}$ then $4$ ($3$ and $\tilde{2}$ respectively) must be present too.

 The summands  $F_{\vec{Y}}$
are given by 
\bea \label{Fwil}
F_{\vec{Y}}=
\prod_{i,j=1}^{2n} \frac{1}{P(u_i,Y_i|u_j,Y_j)}\,.
\eea
Here $	P(a,\lambda|b,\mu)$ is a product of factors obtained from the terms of $X\left(T_{u_i},Y_i|T_{u_j},Y_j\right)$ by usual  
substitution  
\bea
\label{mapX_F}
n T_{\epsilon}^mT_{u_i}T^{-1}_{u_j}\to \left(m \epsilon+u_{ij}\right)^{n}\,,
\eea
\begin{table} 
	\begin{center}
		\resizebox{15cm}{!}{
			\begin{tabular}{| l || c | c | c | c | c | r| }
				\hline
				&$0$ & $1$ & $2$ &$\tilde{2}$&$3$&$4$ \\ \hline\hline
				$0$	&$1$ & $u_{ij}$ & $u_{ij}$ &$u_{ij}$&$u_{ij}$&$u_{ij}(u_{ij}-i)$ \\ \hline
				$1$	&$u_{ij}+i$ & $1$ & $u_{ij}$ &$u_{ij}$&$u_{ij}^2$&$u_{ij}$ \\ \hline
				$2$	&$u_{ij}+i$ & $u_{ij}+i$ & $1$ &$u_{ij}(u_{ij}+i)$&$u_{ij}$&$u_{ij}$ \\ \hline
				$\tilde{2}$	&$u_{ij}+i$ & $u_{ij}+i$ & $u_{ij}(u_{ij}+i)$ &$1$&$u_{ij}$&$u_{ij}$ \\ \hline
				$3$	&$u_{ij}+i$ & $(u_{ij}+i)^2$ & $u_{ij}+i$ &$u_{ij}+i$&$1$&$u_{ij}$ \\ \hline
				$4$	&$(u_{ij}+i)(u_{ij}+2i)$ & $u_{ij}+i$ & $u_{ij}+i$ &$u_{ij}+i$&$u_{ij}+i$&$1$ \\ \hline
			\end{tabular}
		}
	\end{center}
	\caption{This table gives  $P\left(u_i,Y_i|u_j,Y_j\right)$ for all diagrams i.e. $Y_i,Y_j\in \{0,1,2,\tilde{2},3,4\}$. For example $P\left(u_i,2|u_j,1\right)=u_{ij}+i$}
	\label{table:2}
\end{table}
where $m$ and  $n$ are integers. The final result   (alternatively  represented in Table \ref{table:2}) is
\begin{small}
	\bea \label{ZbfWil}
	P(a,\lambda|b,\mu)=
	\begin{cases}
		1       & \, \text{if } \lambda=\mu\\
		\left(a-b+2 \epsilon \right) \left(1+\delta_{4,|\lambda|+|\mu|}\left(-1+a-b+\epsilon(|\lambda|-|\mu|)\right) \right)   & \, \text{if } |\lambda| \ge |\mu|\\
		(a-b) \left(1+\delta_{4,|\lambda|+|\mu|}\left(-1+a-b+2\epsilon+\epsilon(|\lambda|-|\mu|)\right) \right)   & \, \text{if } |\lambda| \le |\mu|
	\end{cases}.\,\,
	\eea
\end{small}
 To derive $P(a,2|b,\tilde{2})$ we can use both the second and third lines of (\ref{ZbfWil}) because they give the same result.
\subsubsection{Example $n=1$}
\label{ex_n1}
\begin{figure}
	\begin{center}
		\begin{tikzpicture}[scale=0.5]
		\draw [rotate around={-45:(13,0)}](12.5,-0.5) rectangle (13.5,0.5);
		\draw[rotate around={-45:(13,0)}] (12.5,0.5) rectangle (13.5,1.5);
		\draw[rotate around={-45:(10.5,0)}] (10,-0.5) rectangle (11,0.5);
		\draw[rotate around={-45:(10.5,0)}] (11,-0.5) rectangle (12,0.5);	
		\draw [rotate around={-45:(5.5,0)}](5,-0.5) rectangle (6,0.5);
		\draw [rotate around={-45:(5.5,0)}]  (5,0.5) rectangle (6,1.5);
		\draw [rotate around={-45:(5.5,0)}] (6,-0.5) rectangle (7,0.5);	
		\draw[rotate around={-45:(3.5,0)}] (3,-0.5) rectangle (4,0.5) ;
		\draw [fill] (-4,0) circle [radius=0.1];
		\draw [rotate around={-45:(-2.5,0)}] (-3,-0.5) rectangle (-2,0.5);
		\draw [rotate around={-45:(-2.5,0)}] (-3,0.5) rectangle (-2,1.5);
		\draw [rotate around={-45:(-2.5,0)}] (-3,-0.5) rectangle (-1,0.5);
		\draw  [rotate around={-45:(-2.5,0)}](-2,0.5) rectangle (-1,1.5);
			\draw (-2.3,-2.5) node{$\{0,4\}$};
				\draw (5,-2.5) node{$\{1,3\}$};
				\draw (12.3,-2.5) node{$\{\tilde{2},2\}$};
		\draw [dotted] (-5,0) -- (17.2,0);
		\draw [dotted] (-5,0.707106) -- (15.2,0.707106);
		\draw [dotted] (-5,-0.707106) -- (15.2,-0.707106);
		\draw (17.9,0) node{$b$};
		\draw (15.9,0.707106) node{$a$};
		\draw (15.9,-0.707106) node{$c$};
		\draw[thick] (-3.5,-0.1) node[scale=2]{,};
		\draw[thick] (0,0) node[scale=2] { $\}$};
		\draw[thick] (-4.5,0) node[scale=2] { $\{$};
			\draw[thick] (4.5,-0.1) node[scale=2]{,};
		\draw[thick] (7.5,0) node[scale=2] { $\}$};
		\draw[thick] (2.5,0) node[scale=2] { $\{$};
		\draw[thick] (11.8,-0.1) node[scale=2]{,};
		\draw[thick] (14.9,0) node[scale=2] { $\}$};
		\draw[thick] (9.5,0) node[scale=2] { $\{$};
		\end{tikzpicture}
	\end{center}
	\caption{The list of allowed diagram pairs for the case when $n=1$ is $\{0,4\}$, $\{1,3\}$, $\{\tilde{2},2\}$, $\{4,0\}$, $\{3,1\}$, $\{2,\tilde{2}\}$.}
	\label{fign1}
\end{figure}
When $n=1$, as  explained in subsection \ref{main},  the dimensions of  the spaces  $u$, $b$, $a$, $c$ are $2$, $2$, $1$, $1$ respectively.  So the sum in  (\ref{PIYoung}) is over the following six pairs $\vec{Y}=\{Y_1,Y_2\}$:
$\{0,4\}$, $\{4,0\}$,  $\{1,3\}$, $\{3,1\}$, $\{2,\tilde{2}\}$, $\{\tilde{2}, 2\}$ (see  Fig.\ref{fign1}).
From  (\ref{Fwil}) and the fact that $P(u_i,Y_i|u_i,Y_i)=1$,   $i=1,2$  (this can be seen from (\ref{ZbfWil}) directly) we will get 
\bea \label{Fwiln1}
F_{\vec{Y}}=\frac{1}{P(u_1,Y_1|u_2,Y_2)P(u_2,Y_2|u_1,Y_1)}\,.
\eea
Below we  derive  $F_{\vec{Y}}$  for all six $\vec{Y}$s.
\begin{enumerate}
	\item When $\vec{Y}=\{0,4\}$ that is $Y_1=0$ and $Y_2=4$ then from (\ref{ZbfWil}) we will obtain
	$P(u_1,Y_1|u_2,Y_2)=u_{12}(u_{12}-2 \epsilon)$ and 
	$P(u_2,Y_2|u_1,Y_1)=(u_{21}+2\epsilon)(u_{21}+4\epsilon)$.
	Inserting these in (\ref{Fwiln1}) we will obtain
	\bea
	F_{\{0,4\}}=\frac{1}{u_{12}(u_{12}-2\epsilon )^2(u_{12}-4\epsilon)}\,.
	\eea
\item When $\vec{Y}=\{4,0\}$ that is $Y_1=4$ and $Y_2=0$ then from (\ref{ZbfWil}) we will obtain
$P(u_1,Y_1|u_2,Y_2)=(u_{12}+2\epsilon)(u_{12}+4\epsilon)$ and
$P(u_2,Y_2|u_1,Y_1)=u_{21}(u_{21}-2\epsilon)$, hence from (\ref{Fwiln1}) we obtain
\bea
F_{\{4,0\}}=\frac{1}{u_{12}(u_{12}+2\epsilon)^2(u_{12}+4\epsilon)}\,.
\eea
\item When $\vec{Y}=\{2,\tilde{2}\}$ that is $Y_1=2$ and $Y_2=\tilde{2}$ then
$P(u_1,Y_1|u_2,Y_2)=(u_{12}+2\epsilon)u_{12}$ and $P(u_2,Y_2|u_1,Y_1)=u_{21}(u_{21}+2\epsilon)$,
so that
\bea
F_{\{2,\tilde{2}\}}=\frac{1}{(u_{12}-2\epsilon)u_{12}^2(u_{12}+2\epsilon)}\,.
\eea	
\item When $\vec{Y}=\{\tilde{2},2\}$ that is $Y_1=\tilde{2}$ and $Y_2=2$ then
$P(u_1,Y_1|u_2,Y_2)=u_{12}(u_{12}+2\epsilon)$ and $P(u_2,Y_2|u_1,Y_1)=(u_{21}+2\epsilon)u_{21}$
and
\bea
F_{\{\tilde{2},2\}}=\frac{1}{(u_{12}-2\epsilon)u_{12}^2(u_{12}+2\epsilon)}\,.
\eea
\item  $\vec{Y}=\{1,3\}$  then
$P(u_1,Y_1|u_2,Y_2)=u_{12}^2$ and $P(u_2,Y_2|u_1,Y_1)=(u_{21}+2\epsilon)^2$
thus
\bea
F_{\{1,3\}}=\frac{1}{(u_{12}-2\epsilon)^2u_{12}^2}\,.
\eea
\item  $\vec{Y}=\{3,1\}$ so
$P(u_1,Y_1|u_2,Y_2)=(u_{12}+2\epsilon)^2$, $P(u_2,Y_2|u_1,Y_1)=u_{21}^2$
and we get
\bea
F_{\{3,1\}}=\frac{1}{u_{12}^2(u_{12}+2\epsilon)^2}\,.
\eea
\end{enumerate}
Inserting  these results in (\ref{PIYoung}) we find 
\bea 
\nonumber
&\Pi_{mat}^{(2)}=
F_{\{0,4\}}+F_{\{4,0\}}+F_{\{2,\tilde{2}\}}+F_{\{\tilde{2},2\}}+F_{\{0,3\}}+F_{\{3,0\}}=
\frac{6}{\left(u_{12}^2+1\right)\left(u_{12}^2+4\right)}\,,
\eea
where the value    $\epsilon=i/2$ is restored. This result is in agreement with the ones obtained 
with other approaches.
\subsubsection{Example $n=2,3,...$}
\label{ex_n2}
Now we consider the case with $n=2$, so our spaces  $u$, $b$, $a$ and $c$ have dimensions $4$, $4$, $2$, $2$ correspondingly. From the constraints explained in section \ref{main}  one can see that there are $90$ possibilities for $\vec{Y}=\{Y_1,Y_2,Y_3,Y_4\}$. Here are some  of them:
 $   \{2,2,\tilde{2}, \tilde{2}\}$, $ \{2,\tilde{2},2,\tilde{2}\}$, $ \{2,\tilde{2},\tilde{2},2\}$, $\{\tilde{2},2,2,\tilde{2}\}$, $\{\tilde{2},2,\tilde{2},2\}$, 
  $\{\tilde{2},\tilde{2},2,2\}$, $ \{3,2,\tilde{2},1\}$, $ \{3,2,1,\tilde{2}\}$, $\{3,\tilde{2},2,1\}$, $\{3,\tilde{2},1,2\}$, $\{3,1,2,\tilde{2}\}$,  ..... .
 For a given $\vec{Y}$ from  (\ref{Fwil}) and (\ref{ZbfWil}) we can derive the corresponding $F_{\vec{Y}}$. \\
 As an example let us derive  $F_{\{2,2,\tilde{2},\tilde{2}\}}$ i.e. $Y_1=2$, $Y_2=2$, $Y_3=\tilde{2}$, $Y_4=\tilde{2}$. From (\ref{ZbfWil})
  one can see straightforwardly that 
 \begin{small}
 	\bea
 	P(u_1,Y_1|u_2,Y_2)=1;\,\,
 	P(u_1,Y_1|u_3,Y_3)=u_{13}(u_{13}+2\epsilon);\,\,
 	P(u_1,Y_1|u_4,Y_4)=u_{14}(u_{14}+2\epsilon);\nonumber\\
 	P(u_2,Y_2|u_1,Y_1)=1;\,\,
 	P(u_2,Y_2|u_3,Y_3)=u_{23}(u_{23}+2\epsilon);\,\,
 	P(u_2,Y_2|u_4,Y_4)=u_{24}(u_{24}+2\epsilon);\nonumber\\
 	P(u_3,Y_3|u_1,Y_1)=u_{31}(u_{31}+2\epsilon);\,\,
 	P(u_3,Y_3|u_2,Y_2)=u_{32}(u_{32}+2\epsilon);\,\,
 	P(u_3,Y_3|u_4,Y_4)=1;\nonumber\\
 	P(u_4,Y_4|u_1,Y_1)=u_{41}(u_{41}+2\epsilon);\,\,
 	P(u_4,Y_4|u_2,Y_2)=u_{42}(u_{42}+2\epsilon);\,\,
 	P(u_4,Y_4|u_3,Y_3)=1.\nonumber
 	\eea
 \end{small}
 After  inserting these in (\ref{Fwil}) we find
 \bea
 F_{\{2,2,\tilde{2},\tilde{2}\}}=\left(u_{13}^2u_{14}^2u_{23}^2u_{24}^2(u_{13}^2-4\epsilon^2)(u_{14}^2-4\epsilon^2)(u_{23}^2-4\epsilon^2)(u_{24}^2-4\epsilon^2)\right)^{-1}\,.
 \eea
   The other $F_{\vec{Y}}$ are derived similarly. From (\ref{PIYoung}) one obtains
   \bea
   \Pi_{mat}^{(4)}=\frac{N^{(4)}}{\prod _{i=1}^4 \prod _{j=1}^{i-1} \left(u_{ij}^2+1\right) \left(u_{ij}^2+4\right)}\,.
   \eea
   where
 \begin{equation}
 \begin{split}
 N^{(4)} =&36 \left(84 e_1^4-57 e_3 e_1^3+19 e_2^2 e_1^2+9 e_3^2 e_1^2-448 e_2 e_1^2+183 e_4 e_1^2+1068 e_1^2-\right.\\
 &-6 e_2^2 e_3 e_1+167 e_2 e_3 e_1-287 e_3 e_1-72 e_3 e_4 e_1+e_2^4-54 e_2^3+693 e_2^2-\\
& \quad \left.
-45 e_3^2+144 e_4^2-2848 e_2+24 e_2^2 e_4-488 e_2 e_4+1148 e_4+3504 \right)\,.
 \end{split}\nonumber
 \end{equation}
 Here the $e_i$ ($i=1,2,3,4$)  are elementary symmetric polynomials in  four variables
 \bea
 e_1=\sum_{i=1}^{4}u_i; \,\,
 e_2=\sum_{1\le i<j\le 3}u_iu_j;\,\,
 e_3=\sum_{1\le i<j<k\le 3}u_iu_ju_k;\,\,
 e_4=u_1 u_2 u_3 u_4\,.
 \eea
 Next let us consider the case when $n=3$  then $\Pi_{mat}^{(6)}$ (\ref {PIYoung}) is given in terms 
 of $1860$ summands. To generate all of them via (\ref{Fwil}) is a mater of few seconds by using for example mathematica.  
 Bringing the $1860$ terms together and canceling out the fake poles is a much more difficult task. The resulting 
 denominator is a very complicated polynomial. Below we present the result for the case when all $u_i$   are very large 
 so the $\epsilon$ in (\ref{ZbfWil}) can be  neglected. We have found
 \bea
 \Pi_{mat}^{(6)}=\frac{N^{(6)}}{\prod _{i=1}^6 \prod _{j=1}^{i-1} u_{ij}^4}+O(\epsilon)
 \eea
 where
 	\bea
	N^{(6)}=216 \left(12 e_6 e_2^3+e_4^2 e_2^2-3 e_3 e_5 e_2^2+75 e_5^2 e_2+e_1 e_4 e_5 e_2-45 e_1 e_3 e_6 e_2-\right.
	\\-126 e_4 e_6 e_2 
 +12 e_4^3-3 e_1 e_3 e_4^2-20 e_1^2 e_5^2+405 e_6^2 +9 e_1 e_3^2 e_5-\nonumber\\ \left.
 -45 e_3 e_4 e_5+81 e_3^2 e_6+75 e_1^2 e_4 e_6-135 e_1 e_5 e_6\right){}^2.\nonumber
 	\eea
Here  $e_i$, $i=1,2,...,6$ are elementary symmetric polynomials from six variables.
In \cite{BFPR3} it was shown that the denominator of $\Pi_{mat}^{(2n)}$ has the following structure
\bea
\Pi_{mat}^{(2n)}(u_1,...,u_{2n})=\frac{N^{(2n)}(u_1,...,u_{2n})}{\prod _{i=1}^{2 n} \prod _{j=1}^{i-1} \left(u_{ij}^2+1\right) \left(u_{ij}^2+4\right)}\,.
\eea
As one can see from (\ref{ZbfWil}) or Table \ref{table:2}, poles of the form $u_{ij}$ cancel  out.

 In order to demonstrate the advantage of our approach  let us compare it with the   method 
     invented in   \cite{BFPR3}.
   If we take into consideration the constraints on the diagrams then for fixed $n$ the number of summands (possible $\vec{Y}$) in (\ref{PIYoung}) is equal to
  \[
  \sum _{k_1=0}^n \sum _{k_2=0}^{n-k_1} \frac{(2 n)!}{\left(k_1!k_2!\left( n-k_1-k_2\right)!\right){}^2}
  \]
  whereas it can be shown that the number of summands when the method of   \cite{BFPR3} is  applied equals to
  \[
  \sum _{k=0}^{n} (2 n)! (2 n-2 k)!
  \]
  For the first few numbers of  scalars these gives Table \ref{comp_with_49}.
  \begin{table}
  	\center
  	\begin{tabular}{ | c || c | c |}
  		\hline
  		&number of summands of  \cite{BFPR3}\ & number of summands of (\ref{PIYoung})\\
  		\hline\hline
  		$n=1$ & 6 & 6 \\
  		\hline
  		$n=2$ & 648 & 90 \\
  		\hline 
  		$n=3$  & 537 840  & 1 860  \\
  		\hline
  		$n=4$  & 1 655 821 440  & 44 730  \\
  		\hline
  	\end{tabular}
  	\caption{}
  		\label{comp_with_49}
  \end{table}
 So we have a good reason to  think that our formula is substantially more efficient for both symbolic and numerical computations.
  
\subsection{Asymptotic factorisation and recursion property of the residue}
Using integral representation (\ref{Pmat1}) it was shown in 
 \cite{BFPR3}  that if one shifts $2k$ of $2n$ rapidities  in $\Pi_{mat}^{(2n)}(u_1,...,u_{2n})$ by  $\Lambda\gg 0$,  then the following factorization formula holds
	\bea
	\label{Pi_fac}
	\Pi_{mat}^{(2n)}(u_1+\Lambda ,\ldots,u_{2k}+\Lambda ,u_{2k+1},\ldots,u_{2n})=\qquad\qquad\qquad\quad \\ \nonumber
=
 \Lambda^{-8k(n-k)}
\left(1+ \frac{4k}{\Lambda }\sum _{m=2k+1}^{2n}u_m
	- 4(n-k)\sum _{i=1}^{2k}\frac{u_i}{\Lambda }+... \right)\times\\
	\times	\Pi_{mat}^{(2k)}(u_1,\ldots,u_{2k})\Pi_{mat}^{2(n-k)}(u_{2k+1},\ldots,u_{2n}).\nonumber
	\eea
 Here we will demonstrate that this property  follows  from   (\ref{PIYoung}) in a straightforward way. For a diagram $Y$ we denote by  $\bar{Y}$ the unique  diagram $Y\ne\bar{Y}$ which has $4-|Y|$ boxes. It is not difficult to see that the arrays of the form
	\bea
	\label{Yl}
	\vec{Y}_{\Lambda}=
	\{
	\underbrace{Y_1,Y_2,...,Y_k,\bar{Y_1},\bar{Y_2},...,\bar{Y_k}}_{2k},\underbrace{Y_{2k+1},Y_{2k+2},...,Y_{n+k},\bar{Y}_{2k+1},\bar{Y}_{2k+2},...,\bar{Y}_{n+k}}_{2n-2k}
	\}
	\eea
 and  those which can be obtained  by  permutations not mixing  the first $2k$ diagrams with the last $2n-2k$ (i.e. those arrays which contain any diagram together with its partner separately in both groups) give the most relevant contributions in above described clustering limit $\Lambda\to \infty$ as seen from Table \ref{table:2}.
  We observe from  (\ref{ZbfWil})  that the both  cases  $\lambda\ne\mu$ and  $\lambda=\mu$ though seem different, in fact  give the same outcome: 
	\begin{small}
		\bea
		\left(\,P(u_i+\Lambda,\lambda|u_n,\mu)P(u_i+\Lambda,\lambda|u_m,\bar{\mu})P(u_n,\mu|u_i+\Lambda,\lambda)P(u_m,\bar{\mu}|u_i+\Lambda,\lambda)\,\right)^{-1}\,\times\\ \nonumber
			\left(P(u_j+\Lambda,\bar{\lambda}|u_n,\mu)P(u_j+\Lambda,\bar{\lambda}|u_m,\bar{\mu})P(u_n,\mu|u_j+\Lambda,\bar{\lambda})P(u_m,\bar{\mu}|u_j+\Lambda,\bar{\lambda})\right)^{-1}=\nonumber\\ \nonumber
		\frac{1}{\Lambda^8}\left(1-\frac{4}{\Lambda}\left(u_{im}+u_{jn}\right)\right)+O\left(\Lambda^{-10}\right)
\,.		\eea
		\end{small}
  Hence it is quite straightforward to observe that contributions of arrays of type  (\ref{Yl}) are given  by
  \bea
  F_{\vec{Y}_{\Lambda}}\sim
  \Lambda^{-8k(n-k)}
  \left(1+ \frac{4k}{\Lambda }\sum _{m=2k+1}^{2n}u_m
  - \frac{4(n-k)}{\Lambda}\sum _{i=1}^{2k}u_i+... \right)\times\\
\,  F_{\{Y_1,Y_2,...,Y_k,\bar{Y}_1,\bar{Y}_2,...,\bar{Y}_k\}}F_{\{Y_{2k+1},Y_{2k+2},...,Y_{n+k},\bar{Y}_{2k+1},\bar{Y}_{2k+2},...,\bar{Y}_{n+k}\}}\,.\,\nonumber
  \eea
  Due to such a nice factorization, summing over all allowed choices of diagrams we arrive  at the desired formula (\ref{Pi_fac}).
  
  In \cite{BFPR3} it was demonstrated that
  \bea
  \label{rec}
  \textit{Res}_{u_2=u_1+2i}\Pi_{mat}^{(2n)}(u_1,\cdots ,u_{2n}) = - \frac{ \Pi_{mat}^{(2n-2)}(u_3,\cdots ,u_{2n})}{2i\displaystyle\prod_{j=3}^{2n}u_{1j}(u_{1j}+i)^2(u_{1j}+2i)}\,.
  \eea
  This result too can be easily  obtained by using our formula for $\Pi^{(2n)}_{mat}$. 
  Indeed the arrays $\vec{Y}$ with nonzero residues  $\textit{Res}_{u_2=u_1+2i}F_{\vec{Y}}\ne0$   must have the structure $\vec{Y}={\{4,0,Y_3,Y_4,...,Y_{2n}\}}$, as can bee seen from Table \ref{table:2}.  From (\ref{Fwil}) 
  \begin{small}
  	\bea
  	& F_{\{4,0,Y_3,Y_4,...,Y_{2n}\}}=
  	\frac{F_{\{Y_3,Y_4,...,Y_{2n}\}}}{P(u_1,4|u_2,0)P(u_2,0|u_1,4)\prod_{j=3}^{2n}P(u_1,4|u_j,Y_j)P(u_2,0|u_j,Y_j)P(u_j,Y_j|u_1,4)P(u_j,Y_j|u_2,0)}\,.\qquad\nonumber
  	\eea
  \end{small}
  Using Table \ref{table:2} for the residue we get
  \bea
   \textit{Res}_{u_2=u_1+2i}F_{\{4,0,Y_3,Y_4,...,Y_{2n}\}}=-
  \frac{F_{\{Y_3,Y_4,...,Y_{2n}\}}}{2i\displaystyle\prod_{j=3}^{2n}u_{1j}(u_{1j}+i)^2(u_{1j}+2i)}\,.
  \eea
 Summing over diagrams (see (\ref{PIYoung})) we recover (\ref{rec}).
\section{Fermions}
\label{fermion}
This section is analogous to the one before, with the only difference that we consider fermions instead of scalars. More precisely, we write down the analogues of ADHM equations that upon localization lead to  (\ref{fPmat1}). As a result, we recover also a combinatorial representation for $\Pi^{(n)}_{mat}$ (see (\ref{Pi_f}) together with (\ref{fpYoung})). By using our formula, we can perform explicit computations and study different properties of $\Pi^{(n)}_{mat}$.
\subsection{Matrix equations and their consequent combinatorial expression}
\label{Fer_mat_eq}
In this section, we will consider the MHV case with no scalars, $N_{\phi}=0$, but equal number of fermions and antifermions $N_{\psi}=N_{\bar{\psi}}=n$:
	\bea
	\label{fPmat1}
&	\Pi^{(n)}_{mat}=\frac{1}{(n!)^3}
	\int_{-\infty}^{\infty}
	\prod_{k=1}^{n}\left(\frac{da_k\, db_k\, dc_k}{(2\pi)^3}\right)
	\frac{\prod^{n}_{i<j}g(a_{ij})g(b_{ij})g(c_{ij})}
	{\prod_{i,j}^{n}f(a_i-b_j)f(c_i-b_j)
		\prod_{i,\alpha}^{n}f(v_{\alpha}-a_i)
		f(\bar{v}_{\alpha}-c_i)}\,,
	\eea
where, like for scalars, we have a multi-integral over the three kinds of nested Bethe Ansatz roots of a $SU(4)$ spin-chain \cite{FPR1, FPR2}, $a_k$, $b_j$ and $c_k$ and the same functions $g(x)$ and $f(x)$ (\ref{f_g_def}) \cite{BSV3,BFPR4}. In this case, too, we want to 
apply localization in order to obtain a combinatorial representation. It turns out that from the 
point of view of localization it is more natural to consider a slightly different quantity 
\bea
\label{Pi_f}
\Pi^{(n)}_{f}=(-)^n\Pi^{(n)}_{mat}\,.
\eea
With simple manipulations one gets
\bea \label{fPmat2}
&	\Pi^{(n)}_{f}=\frac{(-1)^n}{(n!)^3}
\int_{-\infty}^{\infty}
\prod_{k=1}^{n}\left(\frac{da_k\, db_k\, dc_k}{(2\pi i)^3}\right)\times\\ \nonumber
&\times\frac{\prod_{i,j=1}^{'n}a_{ij}(a_{ij}+2\epsilon)b_{ij}
	(b_{ij}+2\epsilon)c_{ij}(c_{ij}+2\epsilon)}{\prod_{i,j}^{n}
	(a_i-b_{j}+\epsilon)(b_{j}-a_i+\epsilon)
	(c_i-b_j+\epsilon)(b_j-c_i+\epsilon)
	\prod_{i,\alpha}^{n}(v_{\alpha}-a_{i})(a_{i}-v_{\alpha}+2\epsilon)
	(\bar{v}_{\alpha}-c_{i})(c_{i}-\bar{v}_{\alpha}+2\epsilon)	}\,.
\eea
 As earlier  $\epsilon\equiv i/2$  and the vanishing factors should be suppressed in the product with a prime.
For this case the  ADHM analogy suggests that  we have to introduce five  vector spaces:
\begin{itemize}
	\item $v$ and $\bar{v}$:  each one  is a  $n$ complex dimensional space associated with the parameters $v_1,v_2,\ldots ,v_{n}$ and
	$\bar{v}_1,\bar{v}_2,\ldots ,\bar{v}_{n}$ respectively
	\item $a$, $b$, and $c$: each one is   a $n$ complex dimensional space associated with the parameters $a_1,a_2,\ldots ,a_{n}$,  $b_1,b_2,\ldots ,b_{n}$ and $c_1,c_2,\ldots ,c_{n}$ 
\end{itemize}
\begin{figure}
	\begin{center}
		\begin{tikzpicture}
		\filldraw [black] 
		(3,0) circle (2pt) 
		(3,2) circle (2pt)
		(0,2) circle (2pt)
		(3,-2) circle (2pt)
		(0,-2) circle (2pt);
		\draw[thick, middlearrow={stealth reversed}] (0,2) .. controls (0.5,2.5) and (2.5,2.5) .. (3,2);
		\draw[ thick, middlearrow={stealth reversed}] (3,2) .. controls (2.5,1.5) and (0.5,1.5) .. (0,2);
		\draw[thick, middlearrow={stealth reversed}] (0,-2) .. controls (0.5,-1.5) and (2.5,-1.5) .. (3,-2);
		\draw[ thick, middlearrow={stealth reversed}] (3,-2) .. controls (2.5,-2.5) and (0.5,-2.5) .. (0,-2);
		\draw[ thick, middlearrow={stealth reversed}] (3,0) .. controls (2.6,0.5) and (2.6,1.5) .. (3,2);
		\draw[thick, middlearrow={stealth reversed}] (3,2) .. controls (3.4,1.5) and (3.4,0.5) .. (3,0);
		\draw[ thick, middlearrow={stealth reversed}] (3,-2) .. controls (3.4,-1.5) and  (3.4,-0.5).. (3,0);
		\draw[thick, middlearrow={stealth reversed}] (3,0) .. controls (2.6,-0.5) and (2.6,-1.5) .. (3,-2);
		\node at (-0.2,2) {$v$};
		\node at (-0.2,-2) {$\bar{v}$};
		\node at (3.3,0) {$b$};
		\node at (3.3,-2) {$c$};
		\node at (3.3,2) {$a$};
		\draw[thick] (1.25,2.7) node[scale=0.8] {$M_{v a}$};
		\draw[thick] (1.25,1.3) node[scale=0.8] {$M_{ a v}$};
		\draw[thick] (1.25,-1.3) node[scale=0.8] {$M_{\bar{v}c}$};
		\draw[thick] (1.25,-2.7) node[scale=0.8] {$M_{ c \bar{v}}$};
		\draw[thick] (2.3,1) node[scale=0.8] {$M_{ba}$};
		\draw[thick] (3.7,1) node[scale=0.8] {$M_{ab}$};
		\draw[thick] (2.3,-1) node[scale=0.8]  {$M_{bc}$};
		\draw[thick] (3.7,-1) node[scale=0.8] {$M_{cb}$};
		\end{tikzpicture}
	\end{center}
	\caption{represents a  quiver diagram where the arrows indicate the linear maps $M_{av}$, $M_{va}$, $M_{ab}$, $M_{ba}$, $M_{bc}$, $M_{cb}$,    $M_{c\bar{v}}$, $M_{\bar{v}c}$ and dots the spaces $v$, $a$, $b$, $c$, $\bar{v}$ on which they act.}
	\label{ferWilsonquiver}
\end{figure}
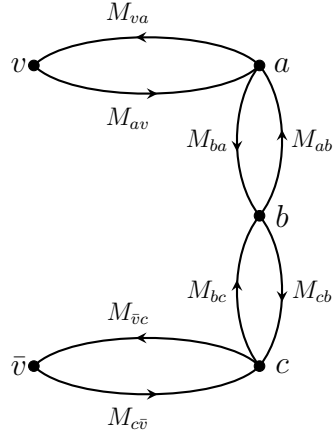
and eight linear maps $M_{v a}$, $M_{va}$, $M_{\bar{v}c}$, $M_{c\bar{v}}$,$M_{ab}$, $M_{ba}$, $M_{cb}$ and  $M_{cb}$ acting among
these spaces as demonstrated in quiver diagram Fig.\ref{ferWilsonquiver}.
The numerator  of (\ref{fPmat2}) is compatible with the following three matrix equations 
\bea \label{fADHMv}
&&M_{ab}M_{ba}+ M_{a\psi}M_{\psi a}=0\;,\\\label{fADHMbarv}
&&M_{cb}M_{bc}+ M_{c\bar{\psi}}M_{\bar{\psi} c}=0\;,\\\label{fADHMu}
&&M_{ba}M_{ab}-M_{bc}M_{cb}=0\,,
\eea
that lead  to consistent final results.
 A suitable \textit{stability condition} is:
There in no proper subspace of $a$ which contains the image $M_{av}$ and is invariant under  $M_{ab}M_{ba}$ and  $M_{ab}M_{bc}M_{cb}M_{ba}$ (i.e.  one acts in all possible ways on image $M_{av}$  the entire space $a$ will be recovered). Similarly there is no proper subspace of $c$ which contains the image $M_{c\bar{v}}$ and is invariant under  $M_{cb}M_{bc}$ and  $M_{cb}M_{ba}M_{ab}M_{bc}$. Finally the direct sum of images of the maps 
$M_{ba}M_{av}$ and $M_{bc}M_{cv}$  covers the space $b$.

We introduce five transformations, three of which $T_a$, $T_b$, $T_c$ are auxiliary while the remaining two     $T_v$, $T_{\bar{v}}$ are diagonal matrices corresponding to  genuine symmetries. 
As before the additional   transformation $T_{\epsilon}\in C^*$ related to the parameter $\epsilon$ is introduced. The denominator of $\Pi^{(n)}_{f}$ (\ref{fPmat2}) dictates the following rule of transformations
\bea
&&M_{ab}\to T_{\epsilon}T_a M_{ab}T_b^{-1}\;, \quad
M_{ba}\to T_{\epsilon} T_b M_{ba}T_a^{-1}\\
&&M_{cb}\to T_{\epsilon} T_c M_{cb} T_b^{-1}\;, \quad
M_{bc}\to T_{\epsilon} T_b M_{bc} T_c^{-1}\\
&&M_{av}\to T_{\epsilon}^2 T_a  M_{av}T_v^{-1}\;, \quad
M_{v a} \to T_v M_{v a}T_a^{-1}\\
&&M_{c\bar{v}}\to T_{\epsilon}^2 T_c  M_{c\bar{v}}T_{\bar{v}}^{-1}\;, \quad
M_{\bar{v} c} \to T_{\bar{v}} M_{\bar{v} c} T_c^{-1}\,.
\eea
It is easy to see that   the  matrix   equations (\ref{fADHMv})-(\ref{fADHMu}) transform in accordance with the numerator of (\ref{fPmat2}).

The  fixed points are given by  the set of matrices 
$M_{ab}$, $M_{ba}$, $M_{bc}$, $M_{cb}$, $M_{av}$, $M_{v a}$,
$M_{c\bar{v}}$ and  $M_{\bar{v} c}$ which are  subject to  equations (\ref{fADHMv})-(\ref{fADHMu}) and are  invariant under   $T_{v}$ , $T_{\bar{v}}$ and $T_{\epsilon}$  up to auxiliary transformations 
 $T_a$, $T_b$ and $T_c$.  One can prove that at fixed points
\bea
M_{v a}=0,\quad
M_{\bar{v} c}=0\,.
\eea
\begin{figure}
	\begin{center}
		\begin{tikzpicture}[scale=0.5]
		\draw [fill] (-1.5,-4.292894) circle [radius=0.1];
		\draw [fill] (-1.5,-10.707106) circle [radius=0.1];
		\draw (-1.5,-7) node{$0_{v}$};
		\draw (3,-7) node{$1_{v}$};
		\draw (7.5,-7) node{$2_{v}$	};
		\draw (12,-7) node{$3_{v}$	};
		\draw (-1.5,-12) node{$0_{\bar{v}}$};
		\draw (3,-12) node{$1_{\bar{v}}$};
		\draw (7.5,-12) node{$2_{\bar{v}}$	};
		\draw (12,-12) node{$3_{\bar{v}}$	};
		\draw [rotate around={-45:(2,-5)}] (1.5,-4.5) rectangle (2.5,-3.5);
		\draw [rotate around={45:(7.5,-5)}] (7,-5.5) rectangle (8,-4.5);
		\draw [rotate around={45:(7.5,-5)}] (7,-4.5) rectangle (8,-3.5);
		\draw [rotate around={45:(12.5,-5)}] (12,-5.5) rectangle (13,-4.5);
		\draw [rotate around={45:(12.5,-5)}] (12,-4.5) rectangle (13,-3.5);
		\draw [rotate around={45:(12.5,-5)}] (12,-6.5) rectangle (13,-5.5);
		\draw[rotate around={-45:(2,-10)}] (2.5,-10.5) rectangle (3.5,-9.5);
		\draw[rotate around={-45:(6.5,-10)}] (7,-10.5) rectangle (8,-9.5);
		\draw[rotate around={-45:(6.5,-10)}] (7,-9.5) rectangle (8,-8.5);
		\draw[rotate around={-45:(10.5,-10)}] (11,-10.5) rectangle (12,-9.5);
		\draw[rotate around={-45:(10.5,-10)}] (11,-9.5) rectangle (12,-8.5);
		\draw[rotate around={-45:(10.5,-10)}] (11,-8.5) rectangle (12,-7.5);
		\draw (17.9,-5) node{$b$};
		\draw (15.9,-4.292894) node{$a$};
		\draw (15.9,-5.707106) node{$c$};
		\draw [thick,dotted] (-2.2,-5) -- (17.2,-5);
		\draw [thick,dotted] (-2.2,-4.292894) -- (15.2,-4.292894);
		\draw [thick,dotted] (-2.2,-5.707106) -- (15.2,-5.707106);
		\draw (17.9,-10) node{$b$};
		\draw (15.9,-9.292894) node{$a$};
		\draw (15.9,-10.707106) node{$c$};
		\draw [thick,dotted] (-2.2,-10) -- (17.2,-10);
		\draw [thick,dotted] (-2.2,-9.292894) -- (15.2,-9.292894);
		\draw [thick,dotted] (-2.2,-10.707106) -- (15.2,-10.707106);
		\end{tikzpicture}
	\end{center}
	\caption{The list of  diagrams for the cases  with fermions and antifermions only. The 
		indices $v$ and $\bar{v}$ indicate weather a diagram starts  from the spaces $a$ or $c$ respectively.
		 As before three dotted lines correspond to the spaces $a$, $b$, $c$.}
	\label{ferm_posibleYD}
\end{figure}
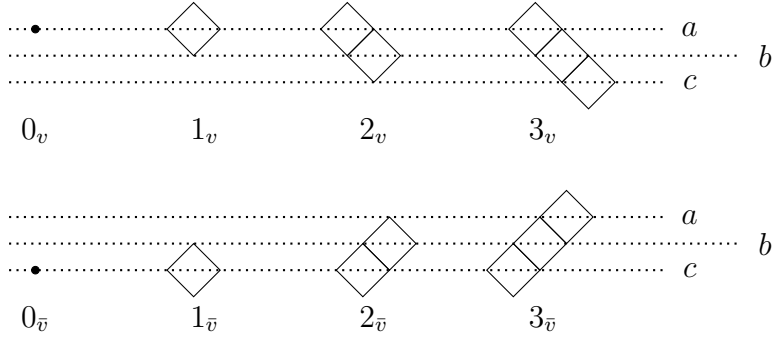
Proceeding as in the case of  scalars we find that the set of admissible 
 diagrams are those  depicted  in Fig.\ref{ferm_posibleYD}.
 
Our final result for fermionic case can be formulated as\footnote{In the next section a more general case,
	 a subcase of which  is the present one,  is considered. The reader can find the expressions for the
	  characters and other details there.} 
\bea \label{fpYoung}
\Pi^{(n)}_{f}=\sum_{\vec{Y}}F_{\vec{Y}}\,,
\eea
where the  sum is  over  collections of $2n$ diagrams  $\vec{Y}\equiv \{Y_1^v,Y_2^v,...,Y_n^v,Y_1^{\bar{v}},Y_2^{\bar{v}},...,Y_n^{\bar{v}}\}$.  $Y^v$ ($Y^{\bar{v}}$) are chosen from the first (second) row of diagrams listed in  Fig.\ref{ferm_posibleYD}. In addition it is required that the total number of boxes  in  upper, middle  and lower lines   (related to the  spaces $a$, $b$ and $c$)    contain  precisely  $n$ boxes each. In terms of Young diagrams the summands in (\ref{fpYoung}) can be represented as
\bea \label{fFwil}
F_{\vec{Y}}=
\prod_{i,j=1}^n
\frac{1}{P(v_i,Y^{v}_i|v_j,Y^{v}_j)P(v_i,Y^{v}_i|\bar{v}_j,Y^{\bar{v}}_j)
	P(\bar{v}_i,Y^{\bar{v}}_i|v_j,Y^{v}_j)P(\bar{v}_i,Y^{\bar{v}}_i|\bar{v}_j,Y^{\bar{v}}_j)}\,.
\eea
\begin{table} 
	\begin{center}
		\resizebox{15cm}{!}{
		\begin{tabular}{| l || c | c | c |c|| c |c| c | r| }
			\hline
			&$0_v$ & $1_v$ & $2_v$ &$3_v$&$0_{\bar{v}}$&$1_{\bar{v}}$ &$2_{\bar{v}}$&$3_{\bar{v}}$\\ \hline\hline
			$0_v$  &$1$ & $v_{ij}$ & $v_{ij}$ &$v_{ij}$&$1$&$1$&$1$&$v_i-\bar{v}_j-i$ \\ \hline
			$1_v$  &$v_{ij}+i$ &$ 1$ & $v_{ij}$ &$v_{ij}$&$1$&$1$ &$v_i-\bar{v}_j$&$1$\\ \hline
			$2_v$  &$v_{ij}+i$ & $v_{ij}+i$ & $1$ &$v_{ij}$&$1$&$v_i-\bar{v}_j+i$ &$1$&$1$\\ \hline
			$3_v$  &$v_{ij}+i$ & $v_{ij}+i$ & $v_{ij}+i$ &$1$&$v_i-\bar{v}_j+2i$&$1$&$1$&$1$ \\ \hline\hline
			$0_{\bar{v}}$	&$1$ & $1$ & $1$ &$\bar{v}_i-v_j-i$&$1$&$\bar{v}_{ij}$&$\bar{v}_{ij}$&$\bar{v}_{ij}$ \\ \hline
			$1_{\bar{v}}$	&$1$ & $1$  &$\bar{v}_i-v_j$&$1$&$\bar{v}_{ij}+i$&$1$&$\bar{v}_{ij}$&$\bar{v}_{ij}$ \\ \hline
			$2_{\bar{v}}$	&$1$ & $\bar{v}_i-v_j+i$ & $1$ &$1$&$\bar{v}_{ij}+i$&$\bar{v}_{ij}+i$&$1$&$\bar{v}_{ij}$ \\ \hline
			$3_{\bar{v}}$	&$\bar{v}_i-v_j+2i$ & $1$ & $1$ &$1$&$\bar{v}_{ij}+i$&$\bar{v}_{ij}+i$&$\bar{v}_{ij}+i$&$1$ \\ \hline
		\end{tabular}
		}
	\end{center}
	\caption{This table lists the factors $P(v_i,Y^{v}_i|v_j,Y^{v}_j)$, $P(v_i,Y^{v}_i|\bar{v}_j,Y^{\bar{v}}_j)$, 
		$P(\bar{v}_i,Y^{\bar{v}}_i|v_j,Y^{v}_j)$ and $P(\bar{v}_i,Y^{\bar{v}}_i|\bar{v}_j,Y^{\bar{v}}_j)$ for 
		 $Y^{v}_i\in \{0,1,2,3\}$ and $Y^{\bar{v}}_i\in \{0_{\bar{v}},1_{\bar{v}},2_{\bar{v}},3_{\bar{v}}\}$. 
		 For example $P(\bar{v}_i,2_{\bar{v}}|\bar{v}_j,1_{\bar{v}})=\bar{v}_{ij}+i$.}
	\label{table:Fermions}
\end{table}
The factors $P(a,\lambda|b,\mu)$ are listed   in Table \ref{table:Fermions}, which  also can be represented by the formula
\begin{small}
	\bea \label{fZbfWil}
	P(a,\lambda^{m}|b,\mu^{n})=
	\begin{cases}
		{\rm when} \quad m=n \quad
		\begin{cases}
			1       & \, \text{if } \lambda=\mu\\
			a-b+i    & \, \text{if } \lambda>\mu\\
			a-b  & \, \text{if } \lambda<\mu
		\end{cases}
		\\
		{\rm when} \quad m\neq n \quad
		\begin{cases}
			1       & \, \text{if } \lambda+\mu\neq 3\\
			a-b+\frac{i}{2}(1+\lambda-\mu)  & \, \text{if } \lambda+\mu=3
		\end{cases}
	\end{cases}
	\eea
\end{small}
where the upper indices  $m$ and $n$ take the values $v$ and $\bar{v}$.
\subsubsection{The cases  $n=1$, $n=2$, $n=3$, ... }
\label{Fer_ex}
If $n=1$ we have four fixed points  given by pairs  of diagrams
$\{0_{v},3_{\bar{v}}\}$, $\{3_{v},0_{\bar{v}}\}$,
$\{1_{v},2_{\bar{v}}\}$ and $\{2_{v},1_{\bar{v}}\}$.
 For the first fixed point   $\vec{Y}=\{0_{v},3_{\bar{v}}\}$, that is  $Y_1^{v}=0_{v}$, $Y_1^{\bar{v}}=3_{\bar{v}}$,
	(\ref{fFwil}) gives
	\begin{small}
		\bea \label{F011}
		F_{\{0_{v},3_{\bar{v}}\}}=
		\frac{1}{P(v_1,0_{v}|v_1,0_{v})P(v_1,0_{v}|\bar{v}_1,3_{\bar{v}})
			P(\bar{v}_1,3_{\bar{v}}|v_1,0_{v})P(\bar{v}_1,3_{\bar{v}}|\bar{v}_1,3_{\bar{v}})}\,.
		\eea
	\end{small}
	From  Table \ref{table:Fermions} 
	\bea
	&&P(v_1,0_{v}|v_1,0_{v})=1\,,\quad
	P(\bar{v}_1,3_{\bar{v}}|\bar{v}_1,3_{\bar{v}})=1\,,\\
	&&P(v_1,0_{v}|\bar{v}_1,3_{\bar{v}})=v_1-\bar{v}_1-i \,,\quad
	P(\bar{v}_1,3_{\bar{v}}|v_1,0_{v})=\bar{v}_1-v_1+2i\,.
	\eea
	Inserting these expressions into (\ref{F011})	we will  find $F_{\{0_{v},3_{\bar{v}}\}}$, the contributions of remaining three fixed points are found similarly: 
	\bea \label{F01103}
	&F_{\{0_{v},3_{\bar{v}}\}}=\frac{1}{(v_1-\bar{v}_1-i)(\bar{v}_1-v_1+2i)}\,,\quad
	F_{\{3_{v},0_{\bar{v}}\}}=\frac{1}{(v_1-\bar{v}_1+2i)(\bar{v}_1-v_1-i)}\,,\\
	&\qquad F_{\{1_{v},2_{\bar{v}}\}}=\frac{1}{(v_1-\bar{v}_1)(\bar{v}_1-v_1+i)}\,,\quad
	F_{\{2_{v},1_{\bar{v}}\}}=\frac{1}{(v_1-\bar{v}_1+i)(\bar{v}_1-v_1)}\,.
	\eea
	Inserting these into (\ref{fpYoung}) and recalling  (\ref{Pi_f}) we obtain
		\bea 
		& \Pi^{(1)}_{mat}=-(F_{\{0_{v},3_{\bar{v}}\}}+F_{\{3_{v},0_{\bar{v}}\}}
		+F_{\{1_{v},2_{\bar{v}}\}}+F_{\{2_{v},1_{\bar{v}}\}})=
		 \frac{4}{(v_1-\bar{v}_1)^2+4}\,,
		\eea	
	which 	coincides with the result in \cite{BFPR4}.
	
		In  \cite{BFPR4} it was shown that $\Pi^{(n)}_{mat}$ can be represented as 
	\bea
	\Pi^{(n)}_{mat}=\frac{N^{(n)}}{\prod_{i<j}^{n}(v_{ij}^2+1)(\bar{v}_{ij}^2+1)\prod_{i,j=1}^{n}((v_i-\bar{v}_j)^2+4)}\,,
	\eea
	where the numerator is a polynomial in $v_i$ and $\bar{v}_i$.
	
	For $n=2$ we have  $28$ fixed points:
	$\{3_v,3_v,0_{\bar{v}},0_{\bar{v}}\}$, $\{2_v,3_v,0_{\bar{v}},1_{\bar{v}}\}$, $\{2_v,2_v,1_{\bar{v}},1_{\bar{v}}\}$, $\{1_v,3_v,0_{\bar{v}},2_{\bar{v}}\}$, $\{1_v,2_v,1_{\bar{v}},2_{\bar{v}}\}$,
	 $\{1_v,1_v,2_{\bar{v}},2_{\bar{v}}\}$, $\{0_v,3_v,0_{\bar{v}},3_{\bar{v}}\}$, $\{0_v,2_v,1_{\bar{v}},3_{\bar{v}}\}$, $\{0_v,1_v,2_{\bar{v}},3_{\bar{v}}\}$, $\{0_v,0_v,3_{\bar{v}},3_{\bar{v}}\}$ 
	 and those obtained by the  permutations of their  first two and last two entries. 
	We used (\ref{fpYoung}) with (\ref{fFwil}) and the  Table \ref{table:Fermions} to derive $\Pi^{(2)}_{mat}$. The result is
			\bea
			N^{(2)}=8 \left(128-18 \bar{s}_1 s_1+26 \bar{s}_2 s_2+2 \bar{s}_1^2 s_1^2-44 \left(\bar{s}_2+s_2\right)+20 \left(\bar{s}_1^2+s_1^2\right)-
			\right.	\\ \left.
			-5 \left(\bar{s}_1^2 s_2+\bar{s}_2 s_1^2\right)-3 \left(\bar{s}_1 \bar{s}_2 s_1+\bar{s}_1 s_1 s_2-\bar{s}_2^2-s_2^2\right)\right)\,,
			\nonumber 
			\eea
where $s_1$ and $s_2$ ($\bar{s}_1$ and $\bar{s}_2$) are the elementary symmetric polynomials in $v_1$ and $v_2$
 ($\bar{v}_1$ and $\bar{v}_2$). 	The result is in full  agreement with \cite{BFPR4}.
		
	For $n=3$ there are  $256$ fixed points. Typical examples are  $\{3_v,3_v,3_v,0_{\bar{v}},0_{\bar{v}},0_{\bar{v}}\}$,\\ $\{2_v,3_v,3_v,0_{\bar{v}},0_{\bar{v}},1_{\bar{v}}\}$. Using mathematica we were able to compute the polynomial $N^{(3)}$ 
	explicitly. Unfortunately the result is too lengthy to be presented here\footnote{The result is available on request.}. 
	 For $n=4$ and $n=5$ we have $2716$ and $31504$ fixed points and mathematica gives the expression as a huge  sum
	  over the fixed points.
\subsubsection{Asymptotic factorisation and recursion property of the residue}	
Let   $\lambda$ be any of the  diagrams $Y^v_1,...,Y^v_k$ and  $\bar{\lambda}$ a diagram  from
 $Y^{\bar{v}}_1,...,Y^{\bar{v}}_k$
having  $3-|\lambda|$ boxes. Similarly  let  $\mu\in \{Y^v_{k+1},...,Y^v_n\}$  and  
$\bar{\mu}\in \{Y^{\bar{v}}_{k+1},...,Y^{\bar{v}}_n\}$ is such that $3-|\mu|$. 
Using Table \ref{table:Fermions} one can check that 
\begin{small}
	\bea\label{PPPP}
\Big\{\,
	P(v_i+\Lambda,\lambda|v_a,\mu) \,  P(v_a,\mu |v_i+\Lambda,\lambda)\,
	P(v_i+\Lambda,\lambda|\bar{v}_b,\overline{\mu})\,  P(\bar{v}_b,\overline{\mu}|v_i+\Lambda,\lambda)\quad \times
	\\ 
	P(\bar{v}_j+\Lambda,\overline{\lambda}|v_a,\mu)  P(v_a,\mu|\bar{v}_j+\Lambda,\overline{\lambda})
	P(\bar{v}_j+\Lambda,\overline{\lambda}|\bar{v}_b,\overline{\mu}) P(\bar{v}_b,\overline{\mu}|\bar{v}_j+\Lambda,\overline{\lambda})
	\Big\}^{-1}= \nonumber\\ \nonumber
	=\frac{1}{\Lambda^4}\left(1-\frac{2}{\Lambda}\left(v_{ia}+\bar{v}_{jb}\right)\right)+O\left(\Lambda^{-6}\right)\,,
	\eea
\end{small}
 where $i,j\in \{1,2,...,k\}$ and $a,b\in \{k+1,k+2,...,n\}$. 
We will denote by $\vec{Y}_{\Lambda}$ a particular choice of $\vec{Y}$  such that for any $i=1,...,k$, $|\vec{Y}^v_i|=3-|\vec{Y}^{\bar{v}}_{p(i)}|$ 
 with $p(i)$ being some permutation of $1,...,k$. It is easy to see that for such a choice of  $\vec{Y}_{\Lambda}$ an analogous property for $k+1,...,n$ holds automatically. So  from (\ref{PPPP}) and (\ref{fFwil}) we get
\begin{small}
	\bea
&	F_{\vec{Y}_{\Lambda}}(v_1+\Lambda,...,v_k+\Lambda,v_{k+1}...,v_n;\bar{v}_1+\Lambda,...,\bar{v}_k+\Lambda,
\bar{v}_{k+1})\sim
\Lambda^{-4k(n-k)}\times \qquad\qquad\\ \nonumber
&	\left(1-\frac{2}{\Lambda}\sum_{i,j=1}^{k}\sum_{a,b=k+1}^{n}\left(v_{ia}+\bar{v}_{jb}\right)\right)
	F_{\{Y^v_1,...,Y^v_k;Y^{\bar{v}}_{1},...,Y^{\bar{v}}_k\}}F_{\{Y^v_{k+1},...,Y^v_n;
		Y^{\bar{v}}_{k+1},...,Y^{\bar{v}}_n\}}\,.
	\eea
\end{small}
One can see from (\ref{fpYoung}) that
	\begin{footnotesize}
		\bea
		&	\Pi^{(n)}_{mat}(v_1+\Lambda,...,v_k+\Lambda,v_{k+1},...,v_n;\bar{v}_1+\Lambda,...,\bar{v}_k+\Lambda,\bar{v}_{k+1},...,\bar{v}_n)	\sim
		\Lambda^{-4k(n-k)}\times
	\qquad\qquad	\\ \nonumber
		&\left(1-\frac{2}{\Lambda}\sum_{i,j=1}^{k}\sum_{a,b=k+1}^{n}\left(v_{ia}+\bar{v}_{jb}\right)\right)
		\Pi^{(k)}_{mat}(v_1,...v_k;\bar{v}_1,...,\bar{v}_k)\Pi^{(n-k)}_{mat}(v_{k+1},...,v_n;\bar{v}_{k+1},...,\bar{v}_n).
		\eea
	\end{footnotesize}

In  \cite{BFPR4} it was demonstrated that
\begin{footnotesize}
	\bea
	\label{recF}
	i\textit{Res}_{\bar{v}_1=v_1+2i}\Pi_{mat}^{(n)}(v_1,\cdots ,v_{n};\bar{v}_1,...,\bar{v}_n) = - \frac{\Pi_{mat}^{(n-1)}(v_2,\cdots ,v_{n};\bar{v}_2,...,\bar{v}_n)}
	{\displaystyle\prod_{j=2}^{n}(v_{1j}+i)v_{1j}(v_1-\bar{v}_j+2i)(v_1-\bar{v}_j+i)}\,.
	\eea
\end{footnotesize}
This result too can be easily  obtained by using our formula for $\Pi^{(n)}_{mat}$. 
Indeed, as can bee seen from Table \ref{table:Fermions}, the arrays $\vec{Y}$ with nonzero residues  $\textit{Res}_{\bar{v}_1=v_1+2i}F_{\vec{Y}}\ne0$    have the structure $\vec{Y}=\{3_v,Y^v_2,...,Y^v_n;0_{\bar{v}},Y^{\bar{v}}_2,...,,Y^{\bar{v}}_n\}$.  From (\ref{fFwil}) we obtain
	\begin{small}
		\bea
	&&	F_{\{3_v,Y^v_2,...,Y^v_n;0_{\bar{v}},Y^{\bar{v}}_2,...,,Y^{\bar{v}}_n\}}=
		\big\{P(v_1,3_v|\bar{v}_1,0_{\bar{v}})P(\bar{v}_1,0_{\bar{v}}|v_1,3_v)
		\times \\
	&&	\prod_{j=2}^{n}P(v_1,3_v|v_j,Y^v_j)P(v_1,3_v|\bar{v}_j,Y^{\bar{v}}_j)
		P(\bar{v}_1,0_{\bar{v}}|v_j,Y^v_j)P(\bar{v}_1,0_{\bar{v}}|\bar{v}_j,Y^{\bar{v}}_j)
		\times \nonumber \\
    &&	P(v_jY^v_j|v_1,3_v)P(v_jY^v_j|\bar{v}_1,0_{\bar{v}})
		P(\bar{v}_j,Y^{\bar{v}}_j|v_1,3_v)P(\bar{v}_j,Y^{\bar{v}}_j|\bar{v}_1,0_{\bar{v}})
		\big\}^{-1}	F_{\{Y^v_2,...,Y^v_n;Y^{\bar{v}}_2,...,,Y^{\bar{v}}_n\}}.
		\nonumber
		\eea
	\end{small}
Using Table \ref{table:Fermions} for the residue we get
\begin{small}
\bea
i\textit{Res}_{\bar{v}_1=v_1+2i}	F_{\{3_v,Y^v_2,...,Y^v_n;0_{\bar{v}},Y^{\bar{v}}_2,...,,Y^{\bar{v}}_n\}}=-
\frac{F_{\{Y^v_2,...,Y^v_n;Y^{\bar{v}}_2,...,,Y^{\bar{v}}_n\}}}
{\displaystyle\prod_{j=2}^{n}(v_{1j}+i)v_{1j}(v_1-\bar{v}_j+2i)(v_1-\bar{v}_j+i)}.
\eea
\end{small}
Summing over the diagrams (see (\ref{fpYoung})) we recover (\ref{recF}).
\section{Young diagram representation for MHV, NMHV and ${\text N}^2$MHV amplitudes}
\label{Gen_formula}
According to \cite{Basso:2015uxa}, inside the hexagonal  Wilson loop in ${\cal N}=4$ SYM the factor accounting for the matrix structure can be again written as a multi-integral over the three kinds of nested Bethe Ansatz roots of a spin-chain $a_i$,  $b_m$, $c_j$, where $i=1,\dots ,K_1$, $m=1,\dots,K_2$, $j=1,\dots,K_3$ \cite{FPR1, FPR2}:
\begin{footnotesize}
	\bea
	\label{gen_Pmat1}
	&	\Pi^{(N_{\psi},N_{\phi},N_{\bar{\psi}})}_{mat}=\frac{1}{K_1!\,K_2!\,K_3!}
	\int_{-\infty}^{\infty}
	\prod_{i=1}^{K_1}\frac{d\,a_i}{2\pi}\prod_{m=1}^{K_2}\frac{d\,b_m}{2\pi}
	\prod_{j=1}^{K_3}\frac{d\,c_j}{2\pi}\times \qquad \qquad\qquad\qquad \qquad\qquad\qquad\qquad\\ \nonumber
	&	\frac{\prod^{K_1}_{i<k}g(a_{ik})\prod^{K_2}_{m<n}g(b_{mn})\prod^{K_3}_{j<k}g(c_{jk})}
	{\prod_{i=1}^{K_1}\prod_{m=1}^{K_2}f(a_i-b_m)\prod_{j=1}^{K_3}\prod_{m=1}^{K_2}f(b_m-c_j)
		\prod_{m=1}^{K_2}\prod_{\ell=1}^{N_{\phi}}f(b_m-u_\ell)\prod_{i=1}^{K_1}\prod_{\alpha=1}^{N_{\psi}}f(a_i-v_{\alpha})
		\prod_{j=1}^{K_3}\prod_{\beta=1}^{N_{\bar{\psi}}}f(c_j-\bar{v}_{\beta})},
	\eea
\end{footnotesize}with the same functions $f(x)$ and $g(x)$ (\ref{f_g_def}). After the usual specification   $\epsilon \equiv\frac{i}{2}$ and shifting the integration variables  $a_i$, $b_m$ and $c_j$  by $\epsilon$ the 
last expression becomes
\bea \label{gen_Pmat2}
&	\Pi^{(N_{\psi},N_{\phi},N_{\bar{\psi}})}_{mat}=\frac{(-1)^{C}}{K_1!K_2!K_3!}
\int_{-\infty}^{\infty}
\prod_{i=1}^{K_1}\frac{d\,a_i}{2\pi i}\prod_{m=1}^{K_2}\frac{d\,b_m}{2\pi i}
\prod_{j=1}^{K_3}\frac{d\,c_j}{2\pi i}\times \qquad  \qquad \qquad \qquad \qquad \qquad\\ \nonumber
& \frac{\prod_{i,k=1}^{'K_1}a_{ik}(a_{ik}+2\epsilon)\prod_{m,n=1}^{'K_2}b_{mn}
	(b_{mn}+2\epsilon)\prod_{j,k=1}^{'K_3}c_{jk}(c_{jk}+2\epsilon)}{\prod_{m=1}^{K_2}\prod_{i=1}^{K_1}
	(a_i-b_{m}+\epsilon)(b_{m}-a_i+\epsilon)
	\prod_{m=1}^{K_2}\prod_{j=1}^{K_3}(c_j-b_{m}+\epsilon)(b_{m}-c_j+\epsilon)
}\times  \qquad \qquad \qquad \qquad\, \\ \nonumber
& \frac{1}{
	\prod_{m=1}^{K_2}\prod_{\ell=1}^{N_{\phi}}(u_{\ell}-b_{m})(b_{m}-u_{\ell}+2\epsilon)
	\prod_{i=1}^{K_1}\prod_{\alpha=1}^{N_{\psi}}(v_{\alpha}-a_{i})(a_{i}-v_{\alpha}+2\epsilon)
	\prod_{j=1}^{K_3}\prod_{\beta=1}^{N_{\bar{\psi}}}(\bar{v}_{\beta}-c_{j})(c_{j}-\bar{v}_{\beta}+2\epsilon)}\,,
\eea
where $C=K_1K_2+K_2K_3+K_2N_{\phi}+K_1N_{\psi}+K_3N_{\bar{\psi}}$,  the prime  on the product symbol again  indicates that all the vanishing factors  must be ignored. Here $u_{\ell}$ are the rapidities of the scalars and $v_{\alpha}$ and $\bar{v}_{\beta}$ are the rapidities of the 
fermions and anti-fermions, respectively. $N_{\phi}$, $N_{\psi}$ and $N_{\bar{\psi}}$ are the numbers of scalars, fermions and anti-fermions. $K_{j}$ $j=1,2,3$ satisfy the following conditions (see \cite{Basso:2015uxa})
\bea
\label{res1}
&N_{\psi}-2K_{1}+K_{2}=\delta_{r_b,3}\,,\qquad\\
\label{res2}
&N_{\bar{\psi}}-2K_{3}+K_{2}=\delta_{r_b,1}\,,\qquad\\
\label{res3}
&\,N_{\phi}+K_{1}-2K_{2}+K_3=\delta_{r_b,2}\,,
\eea
where $0 \leq r_b\leq 4$  is the $R$ charge carried by the bottom pentagon. For ${\rm MHV}$ and ${\rm N^2MHV}$ the parameter $r_b$
 is fixed  and takes the values  $r_b=0$ and $r_b=4$ respectively. In NMHV case $r_b$ may assume any of the  five allowed  values. 

  The  MHV case with $2K_1=2K_3=K_2=2n$, $N_{\psi}=N_{\bar{\psi}}=0$, $N_{\phi}=2n$ is considered in grate details in section \ref{scalars} while the  MHV  amplitudes with $K_1=K_2=K_3=n$, $N_{\psi}=N_{\bar{\psi}}=n$, $N_{\phi}=0$ are considered in  section \ref{fermion}. Here we treat the general case.
\subsection{Matrix equations}
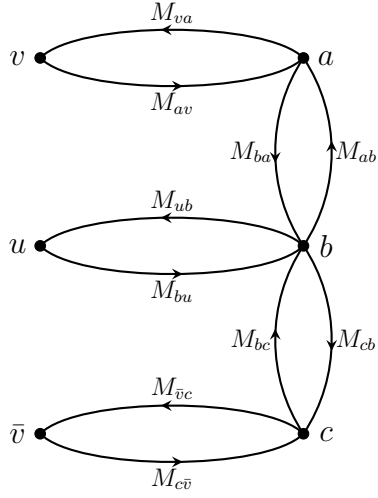
\begin{figure}
	\begin{center}
		\begin{tikzpicture}
		\filldraw [black] (-0.5,0) circle (2pt)
		(3,0) circle (2pt) 
		(3,2.5) circle (2pt)
		(-0.5,2.5) circle (2pt)
		(3,-2.5) circle (2pt)
		(-0.5,-2.5) circle (2pt);
		\draw[thick, middlearrow={stealth reversed}] (-0.5,0) .. controls (0,0.5) and (2.5,0.5) .. (3,0);
		\draw[ thick, middlearrow={stealth reversed}] (3,0) .. controls (2.5,-0.5) and (0,-0.5) .. (-0.5,0);
		\draw[thick, middlearrow={stealth reversed}] (-0.5,2.5) .. controls (0,3) and (2.5,3) .. (3,2.5);
		\draw[ thick, middlearrow={stealth reversed}] (3,2.5) .. controls (2.5,2) and (0,2) .. (-0.5,2.5);
		\draw[thick, middlearrow={stealth reversed}] (-0.5,-2.5) .. controls (0,-2) and (2.5,-2) .. (3,-2.5);
		\draw[ thick, middlearrow={stealth reversed}] (3,-2.5) .. controls (2.5,-3) and (0,-3) .. (-0.5,-2.5);
		\draw[ thick, middlearrow={stealth reversed}] (3,0) .. controls (2.5,0.75) and (2.5,1.75) .. (3,2.5);
		\draw[thick, middlearrow={stealth reversed}] (3,2.5) .. controls (3.5,1.75) and (3.5,0.75) .. (3,0);
		\draw[ thick, middlearrow={stealth reversed}] (3,-2.5) .. controls (3.5,-1.75) and  (3.5,-0.75).. (3,0);
		\draw[thick, middlearrow={stealth reversed}] (3,0) .. controls (2.5,-0.75) and (2.5,-1.75) .. (3,-2.5);
		\node at (-0.8,0) {$u$};
		\node at (-0.8,2.5) {$v$};
		\node at (-0.8,-2.5) {$\bar{v}$};
		\node at (3.3,0) {$b$};
		\node at (3.3,-2.5) {$c$};
		\node at (3.3,2.5) {$a$};
		\draw[thick] (1.25,0.6) node[scale=0.8] {$M_{u b}$};
		\draw[thick] (1.25,-0.6) node[scale=0.8] {$M_{ b u}$};
		\draw[thick] (1.25,3.1) node[scale=0.8] {$M_{v a}$};
		\draw[thick] (1.25,1.9) node[scale=0.8] {$M_{ a v}$};
		\draw[thick] (1.25,-1.9) node[scale=0.8] {$M_{\bar{v}c}$};
		\draw[thick] (1.25,-3.1) node[scale=0.8] {$M_{ c \bar{v}}$};
		\draw[thick] (2.3,1.25) node[scale=0.8] {$M_{ba}$};
		\draw[thick] (3.7,1.25) node[scale=0.8] {$M_{ab}$};
		\draw[thick] (2.3,-1.25) node[scale=0.8]  {$M_{bc}$};
		\draw[thick] (3.7,-1.25) node[scale=0.8] {$M_{cb}$};
		\end{tikzpicture}
	\end{center}
	\caption{represents a  quiver diagram where the arrows indicate the linear maps 
		$M_{av}$, $M_{va}$, $M_{ab}$, $M_{ba}$, $M_{bu}$, $M_{ub}$, $M_{bc}$, $M_{cb}$,   
		 $M_{c\bar{v}}$, $M_{\bar{v}c}$ and dots the 
		spaces $v$, $a$, $u$, $b$, $c$, $\bar{v}$ on which they act.}
	\label{Wilsonquiver}
\end{figure}
Let us start by constructing a more general system of matrix equations  corresponding to the integral representation of $\Pi^{(N_{\psi}N_{\phi}N_{\bar{\psi}})}_{mat}$
 (\ref{gen_Pmat2}). As a result, by means of localization, we will get an efficient    combinatorial procedure for the evaluation of these integrals. Similar to the case considered in 
 section \ref{fermion} here too it is more natural to consider a slightly different quantity
 defined as
 \bea
 \label{piF_piMAT}
 \Pi^{(N_{\psi}N_{\phi}N_{\bar{\psi}})}_{f}=
 (-)^{m}\Pi^{(N_{\psi}N_{\phi}N_{\bar{\psi}})}_{mat}\,, 
 \eea
 where
 \bea
 \label{m}
 m=K_1K_2+K_2K_3+K_2N_{\phi}+K_1N_{\psi}+K_3N_{\bar{\psi}}+K_1+K_2+K_3\,.
 \eea  
 As the integrand of   (\ref{gen_Pmat2}) suggests, 
   one needs  to introduce six spaces:
\begin{itemize}
	\item $u$, 	$v$, $\bar{v}_1$   that are  $N_{\phi}$, $N_{\psi}$ and $N_{\bar{\psi}}$ dimensional complex vector  spaces  respectively. They are   connected to parameters
	$u_1, u_2,...,u_{N_{\phi}}$, 	$v_1,v_2,...,v_{N_{\psi}}$ 	and	$\bar{v}_1, \bar{v}_2,...,\bar{v}_{N_{\bar{\psi}}}$
	\item $a$, $b$ and $c$ which are $K_1$, $K_2$ and $K_3$ dimensional complex vector spaces  related to
	$a_1, a_2, ...,a_{K_1}$, 	$b_1, b_2, ..., b_{K_2}$ and $c_1, c_2, ...,c_{K_3}$
\end{itemize}
and ten linear maps (matrices)  $M_{ab}$, $M_{ba}$, $M_{bc}$, $M_{cb}$,  $M_{bu}$, $M_{ub}$, $M_{av}$, $M_{v a}$, $M_{c\bar{v}}$, $M_{\bar{v} c}$  acting among  these spaces as indicated by the quiver diagram given  in Fig.\ref{Wilsonquiver}.

 Admissible matrix equations 
 compatible with  the form of the  numerator in  (\ref{gen_Pmat2}) can be chosen as:
\bea \label{ADHMv}
&M_{ab}M_{ba}+ M_{av}M_{v a}=0\;,\\\label{ADHMbarv}
&M_{cb}M_{bc}+ M_{c\bar{v}}M_{\bar{v} c}=0\;,\\\label{ADHMu}
&M_{ba}M_{ab}-M_{bc}M_{cb}+M_{bu}M_{u b}=0\,.
\eea
 Similar to previous cases we introduce  auxiliary transformations    $T_a$, $T_b$, $T_c$  in addition with  three   diagonal matrices   $T_u$, $T_v$, $T_{\bar{v}}$  corresponding to  genuine symmetries. 
The role of   $T_{\epsilon}\in C^{*}$  is the same as in previous cases. The denominator of  (\ref{gen_Pmat2}) dictates the  transformation rules of the matrices $M$:
\bea
\label{tr_first}
&&M_{ab}\to T_{\epsilon}T_a M_{ab}T_b^{-1}\;, \quad
M_{ba}\to T_{\epsilon} T_b M_{ba}T_a^{-1}\\
&&M_{cb}\to T_{\epsilon} T_c M_{cb} T_b^{-1}\;, \quad
M_{bc}\to T_{\epsilon} T_b M_{bc} T_c^{-1}\\
&&M_{bu}\to T_{\epsilon}^2 T_b M_{bu}T_u^{-1}\;, \quad
M_{u b} \to T_uM_{u b} T_b^{-1}\\
&&M_{av}\to T_{\epsilon}^2 T_a  M_{av}T_v^{-1}\;, \quad
M_{v a} \to T_v M_{v a}T_a^{-1}\\
&&M_{c\bar{v}}\to T_{\epsilon}^2 T_c  M_{c\bar{v}}T_{\bar{v}}^{-1}\;, \quad
M_{\bar{v} c} \to T_{\bar{v}} M_{\bar{v} c} T_c^{-1}\,.
\label{tr_last}
\eea
Under these  transformations  
the left hand sides of the  matrix   equations (\ref{ADHMv})-(\ref{ADHMu}) transform respectively as: 
\bea \label{gen_TAADHMa}
&M_{ab}M_{ba}+ M_{av}M_{v a}\to 
T_{\epsilon}^2  T_a (M_{ab}M_{ba}+ M_{av}M_{v a}) T_a^{-1}\;,\\\label{gen_TCADHMc}
&M_{cb}M_{bc}+ M_{c\bar{v}}M_{\bar{v} c}\to 
T_{\epsilon}^2 T_c  (M_{cb}M_{bc}+ M_{c\bar{v}}M_{\bar{v} c}) T_c^{-1}\;,\\\label{gen_TBADHMd}
&M_{ba}M_{ab}-M_{bc}M_{cb}+M_{bu }M_{u b}\to
T_{\epsilon}^2 T_b (M_{ba}M_{ab}-M_{bc}M_{cb}+M_{bu}M_{ub}) T_b^{-1}
\eea
in accordance  with the form of numerator of (\ref{gen_Pmat2}).
\subsection{The fixed points of the moduli space}
\textit{Moduli space}: by definition a point in moduli space ${\cal M}_n$ is a set of matrices \\
$\{ M_{ab},M_{ba},M_{bc},M_{cb},M_{bu},M_{u b},M_{av},M_{va},M_{c\bar{v}},M_{\bar{v} c}\}$ satisfying
 the stability condition and the equations (\ref{ADHMv})-(\ref{ADHMu}) modulo the  equivalence relation 
\bea
&& \{ M_{ab},M_{ba},M_{bc},M_{cb},M_{bu},M_{u b},M_{av},M_{va},M_{c\bar{v}},M_{\bar{v} c}\} \sim
\{ T_a M_{ab} T_b^{-1},  T_b M_{ba} T_a^{-1}, \qquad\\\nonumber
&& \qquad\qquad T_b M_{bc} T_c^{-1}, T_c M_{cb} T_b^{-1},T_b M_{bu}, M_{u b} T_b^{-1},T_a M_{av},M_{va}T_a^{-1},T_cM_{c\bar{v}},M_{\bar{v} c}T_c^{-1}\}.
\eea
 In addition 
we  supplement the matrix equations with a stability condition. 
Roughly speaking   this condition states that starting with the images of $v$, $u$, $\bar{v}$ in $a$, $b$,  $c$
 respectively  and  acting by matrices   $M_{ba}$, $M_{ab}$, $M_{cb}$, $M_{bc}$ 
in various consistent ways one completely covers the each of spaces  $a$, $b$ and  $c$.

Our next step is to find the points of the moduli space fixed under transformations 
$T_{\epsilon}$, $T_{v}$,   $T_{u}$, $T_{\bar{v}}$. 
 It is possible to show that at fixed points 
\bea
M_{u b} =0,\quad
M_{v a}=0,\quad
M_{\bar{v} c}=0\,.
\eea
\begin{figure}
	\begin{center}
		\begin{tikzpicture}[scale=0.5]
		\draw [fill] (-1.5,0) circle [radius=0.1];
		\draw [fill] (-1.5,-4.292894) circle [radius=0.1];
		\draw [fill] (-1.5,-10.707106) circle [radius=0.1];
		\draw (-1.5,-7) node{$0_{u}$};
		\draw (1.5,-7) node{$1_{u}$};
		\draw (4.5,-7) node{$2_{u}$};
		\draw (7.5,-7) node{$\tilde{2}_{u}$	};
		\draw (10.5,-7) node{$3_{u}$	};
		\draw (13.5,-7) node{$4_{u}$	};
		\draw (-1.5,-2) node{$0_{v}$};
		\draw (3,-2) node{$1_{v}$};
		\draw (7.5,-2) node{$2_{v}$	};
		\draw (12,-2) node{$3_{v}$	};
		\draw (-1.5,-12) node{$0_{\bar{v}}$};
		\draw (3,-12) node{$1_{\bar{v}}$};
		\draw (7.5,-12) node{$2_{\bar{v}}$	};
		\draw (12,-12) node{$3_{\bar{v}}$	};
		\draw[rotate around={-45:(1.5,-5)}] (1,-5.5) rectangle (2,-4.5) ;
		\draw [rotate around={-45:(4.5,-5)}](4,-5.5) rectangle (5,-4.5);
		\draw[rotate around={-45:(4.5,-5)}] (4,-4.5) rectangle (5,-3.5);
		\draw[rotate around={-45:(7.5,-5)}] (7,-5.5) rectangle (8,-4.5);
		\draw[rotate around={-45:(7.5,-5)}] (8,-5.5) rectangle (9,-4.5);	
		\draw [rotate around={-45:(10.5,-5)}](10,-5.5) rectangle (11,-4.5);
		\draw [rotate around={-45:(10.5,-5)}]  (10,-4.5) rectangle (11,-3.5);
		\draw [rotate around={-45:(10.5,-5)}] (11,-5.5) rectangle (12,-4.5);	
		\draw [rotate around={-45:(13.5,-5)}] (13,-5.5) rectangle (14,-4.5);
		\draw [rotate around={-45:(13.5,-5)}] (13,-4.5) rectangle (14,-3.5);
		\draw [rotate around={-45:(13.5,-5)}] (14,-5.5) rectangle (15,-4.5);
		\draw  [rotate around={-45:(13.5,-5)}](14,-4.5) rectangle (15,-3.5);	
		\draw [rotate around={-45:(2,0)}] (1.5,0.5) rectangle (2.5,1.5);
		\draw [rotate around={45:(7.5,0)}] (7,-0.5) rectangle (8,0.5);
		\draw [rotate around={45:(7.5,0)}] (7,0.5) rectangle (8,1.5);
		\draw [rotate around={45:(12.5,0)}] (12,-0.5) rectangle (13,0.5);
		\draw [rotate around={45:(12.5,0)}] (12,0.5) rectangle (13,1.5);
		\draw [rotate around={45:(12.5,0)}] (12,-1.5) rectangle (13,-0.5);
		\draw[rotate around={-45:(2,-10)}] (2.5,-10.5) rectangle (3.5,-9.5);
		\draw[rotate around={-45:(6.5,-10)}] (7,-10.5) rectangle (8,-9.5);
		\draw[rotate around={-45:(6.5,-10)}] (7,-9.5) rectangle (8,-8.5);
		\draw[rotate around={-45:(10.5,-10)}] (11,-10.5) rectangle (12,-9.5);
		\draw[rotate around={-45:(10.5,-10)}] (11,-9.5) rectangle (12,-8.5);
		\draw[rotate around={-45:(10.5,-10)}] (11,-8.5) rectangle (12,-7.5);
		\draw [thick,dotted] (-2.2,0) -- (17.2,0);
		\draw [thick,dotted] (-2.2,0.707106) -- (15.2,0.707106);
		\draw [thick,dotted] (-2.2,-0.707106) -- (15.2,-0.707106);
		\draw (17.9,0) node{$b$};
		\draw (15.9,0.707106) node{$a$};
		\draw (15.9,-0.707106) node{$c$};
		\draw (17.9,-5) node{$b$};
		\draw (15.9,-4.292894) node{$a$};
		\draw (15.9,-5.707106) node{$c$};
		\draw [thick,dotted] (-2.2,-5) -- (17.2,-5);
		\draw [thick,dotted] (-2.2,-4.292894) -- (15.2,-4.292894);
		\draw [thick,dotted] (-2.2,-5.707106) -- (15.2,-5.707106);
		\draw (17.9,-10) node{$b$};
		\draw (15.9,-9.292894) node{$a$};
		\draw (15.9,-10.707106) node{$c$};
		\draw [thick,dotted] (-2.2,-10) -- (17.2,-10);
		\draw [thick,dotted] (-2.2,-9.292894) -- (15.2,-9.292894);
		\draw [thick,dotted] (-2.2,-10.707106) -- (15.2,-10.707106);
		\end{tikzpicture}
	\end{center}
		\caption{The list of  diagrams if one has  fermions  antifermions and scalars. The diagrams are labeled by $0_v$, ..., $3_v$, $0_u$, ..., $4_u$,  $0_{\bar{v}}, ...,3_{\bar{v}}$, where the  indicate coincides with the number  of boxes in a diagram and the indices $v$, $u$ and $\bar{v}$ tell weather a diagram is connected to fermions scalars or antifermions. As before three dotted lines correspond to the spaces $a$, $b$, $c$.}
	\label{posibleYD_gen}
\end{figure}
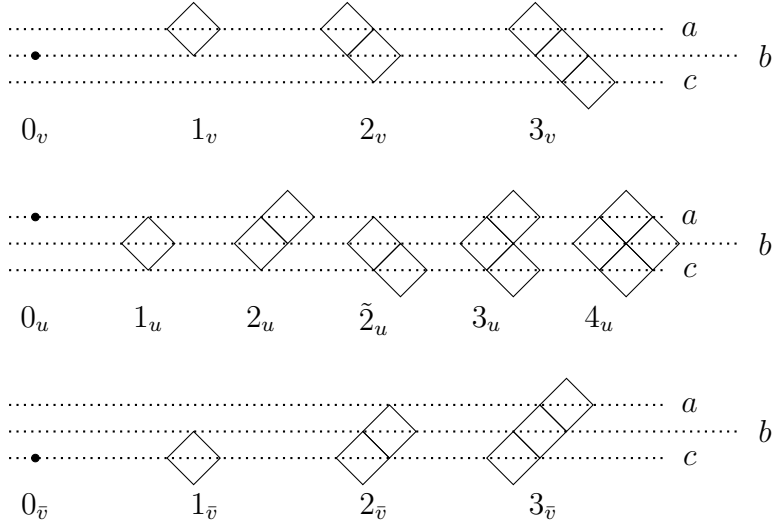
Construction of fixed points goes parallel to the cases discussed   in previous cases (see section \ref{possDIAG}). 
In fact the general case under consideration is a simple combination of the purely bosonic and the case with several fermion antifermion pairs. The complete set of allowed diagrams is depicted in  Fig.\ref{posibleYD_gen}.
 Remind that  a box of a diagram located on the dotted lines  $a$, $b$ or $c$  represents a basis vector in the respective space. 

A fixed point is represented by $N_{\psi}+N_{\phi}+N_{\bar{\psi}}$ diagrams from  Fig.\ref{posibleYD_gen}.
Since  the dimension of $u$ is $N_{\phi}$ we must have $N_{\phi}$  diagrams   
from the second row.
The dimension of $v$ and $\bar{v}$ are  $N_{\psi}$ and $N_{\bar{\psi}}$  respectively hence  we need  $N_{\psi}$   diagrams   from the first and $N_{\bar{\psi}}$ diagrams from the third row.
In addition  to match the dimensions of spaces  $a$, $b$ and  $c$   we should have in total 
 $K_1$ boxes on first dotted line,  $K_2$ boxes on the  second and $K_3$ on the third lines.
 
To  apply localisation  we  need  the  tangent space of ${\cal M}$ at the fixed points. As usual we start with the 
  total unconstrained space   $\delta M$ then  ``subtract" subspaces corresponding  to equations 
  (\ref{ADHMv})-(\ref{ADHMu}) and auxiliary transformations.
  
Transformation laws (\ref{tr_first})-(\ref{tr_last}) dictate the following structure of the variations of matrices $M$
\bea
\delta M_{ab}\in T_{\epsilon} a \otimes b^{*}, \quad
\delta M_{ba}\in T_{\epsilon} b \otimes a^{*},\quad
\delta M_{cb}\in T_{\epsilon} c \otimes b^{*},\quad
\delta M_{bc}\in T_{\epsilon} b \otimes c^{*},\\ \nonumber
\delta M_{a v}\in T_{\epsilon}^2 a \otimes v^{*}, \quad
\delta M_{v a}\in  v \otimes a^{*},\quad
\delta M_{c \bar{v}}\in T_{\epsilon}^2 c \otimes \bar{v}^{*}, \quad
\delta M_{\bar{v} c}\in \bar{v} \otimes c^{*},\\ \nonumber
\delta M_{b u}\in T_{\epsilon}^2 b \otimes u^{*}, \quad
\delta M_{u b}\in u \otimes b^{*}.
\eea
Subtracting from the total space of unconstrained  deformations
 the spaces  (\ref{eq_sp}) and   (\ref{aux_sp}), corresponding to equations and  auxiliary transformations,
for the tangent space we get 
\bea \label{gen_wilchar2}
&\chi_{\vec{Y}}= T_{\epsilon}\left((a+c) b^{*} +b(a^{*}+c^{*})\right)+T_{\epsilon}^2 b u^{*}
+u  b^{*}+\\ \nonumber
&+T_{\epsilon}^2 av^{*}+v a^{*}
+T_{\epsilon}^2 c\bar{v}^{*}+\bar{v}c^{*}
-(T_{\epsilon}^2+1)\left(aa^{*} + bb^{*}+cc^{*}\right)\,.
\eea
The characters  of the spaces $v$, $u$ and $\bar{v}$ are 
 (the same latter is used  for both the space and its character)
\bea
v=\sum_{i=1}^{N_{\psi}}T_{v_i} \,,\quad
u=\sum_{i=1}^{N_{\phi}}T_{u_i} \,,\quad
\bar{v}=\sum_{i=1}^{N_{\bar{\psi}}}T_{\bar{v}_i} \,.\quad
\eea
The dual characters  $u^{*}$, $v^{*}$, $\bar{v}^{*}$ are obtained by substitution 
 $T_{u_i}\to T^{-1}_{u_i}$, $T_{v_i}\to T^{-1}_{v_i}$ and $T_{\bar{v}_i}\to T^{-1}_{\bar{v}_i}$.
  Similarly  for the spaces $a$, $b$ and $c$ we have
\bea
\label{abc}
a=\sum_{k}a_k\,,\quad
b=\sum_{k}b_k\,,\quad
c=\sum_{k}c_k
\eea
and the conjugates are obtained by replacing the summands by their inverses.
Examining the structure of the diagrams we get convinced that the summands in equations (\ref{abc})
explicitly are given by
\bea\label{a}
a_k=
\begin{cases}
	T_{\epsilon} T_{u_k}      & \quad \text{if } Y=2_{u},3_{u},4_{u}\\
	0  & \quad \text{if } Y=0_{u},1_{u},\tilde{2}_{u},0_{v},0_{\bar{v}},1_{\bar{v}},2_{\bar{v}},\\
	T_{v_k}      & \quad  \text{if } Y=1_{v},2_{v},3_{v}\\
	T_{\epsilon}^2	T_{\bar{v}_k}       & \quad  \text{if } Y=3_{\bar{v}}
\end{cases}
\,,
\eea
\bea\label{c}
c_k=
\begin{cases}
	T_{\epsilon} T_{u_k}      & \quad  \text{if } Y=\tilde{2}_{u},3_{u},4_{u}\\
	0  & \quad \text{if } Y=0_{u},1_{u},2_{u},0_{v},1_{v},2_{v},0_{\bar{v}}\\
	T_{\epsilon}^2	 T_{v_k}       & \quad \text{if } Y=3_{v}\\
	T_{\bar{v}_k}     & \quad  \text{if } Y=1_{\bar{v}},2_{\bar{v}},3_{\bar{v}}
\end{cases}
\,,
\eea
\bea
\label{b}
b_k=
\begin{cases}
	T_{u_{k}}       & \quad \text{if } Y=1_{u},2_{u},\tilde{2}_{u},3_{u}\\
	(T_{\epsilon}^2+1)T_{u_{k}}         & \quad \text{if } Y=4_{u}\\
	0  & \quad \text{if } Y=0_{u},0_{v},1_{v},0_{\bar{v}},1_{\bar{v}}\\
	T_{\epsilon}	 T_{v_{k}}       & \quad \text{if } Y=2_{v},3_{v}\\
	T_{\epsilon}	T_{\bar{v}_{k}}     & \quad \text{if } Y=2_{\bar{v}},3_{\bar{v}}
\end{cases}
\eea
In view of above decompositions the detailed   structure of the  character (\ref{gen_wilchar2}) takes the form 
\bea
\nonumber
\chi_{\vec{Y}}=\sum_{i,j=1}^{N_{\psi}}\sum_{m,n=1}^{N_{\phi}}\sum_{k,l=1}^{N_{\bar{\psi}}}
\Big( X({T_{v_i}},Y^{v}_i|T_{v_j},Y^{v}_j)+X(T_{v_i},Y^{v}_i|T_{u_n},Y^{u}_n)+X(T_{v_i},Y^{v}_i|T_{\bar{v}_k},Y^{\bar{v}}_k)+
\\ \nonumber
+ X(T_{u_n},Y^u_n|T_{v_i},Y^{v}_i)+ X(T_{u_n},Y^u_n|T_{u_m},Y^u_m)	+ X(T_{u_n},Y^u_n|T_{\bar{v}_k},Y^{\bar{v}}_k)+\qquad
\\ \label{gen_X}
+	X(T_{\bar{v}_k},Y^{\bar{v}}_k|T_{v_i},Y^{v}_i)+X(T_{\bar{v}_k},Y^{\bar{v}}_k|T_{u_n},Y^u_n)+
X(T_{\bar{v}_k},Y^{\bar{v}}_k|T_{\bar{v}_l},Y^{\bar{v}}_l)\Big),\qquad\quad
\eea
where the summands, derived  from (\ref{gen_wilchar2}) and (\ref{a})-(\ref{b}), are listed in Table \ref{gen_CH_table}.
\begin{table}
	\begin{center}
			\resizebox{15.5cm}{!}{
			\begin{tabular}{l|cccc||cccccc||cccc}
			&$0_{v}$&$1_{v}$&$2_{v}$&$3_{v}$&$0_{u}$&$1_{u}$&$2_{u}$&$\tilde{2}_{u}$&$3_{u}$&$4_{u}$
			&$0_{\bar{v}}$&$1_{\bar{v}}$&$2_{\bar{v}}$&$3_{\bar{v}}$\\
			\hline	
			$0_{v}$&$	0 $&$ 1 $&$ 1$ &$ 1$ &$ 0$ &$ 0$ &$ T_{\epsilon}^{-1} $&$ 0$ &$ T_{\epsilon}^{-1}$
			&$ T_{\epsilon}^{-1}$ &$ 0$ &$ 0$ &$ 0$ &$ T_{\epsilon}^{-2}$ \\
			$1_{v}$&	$T_{\epsilon}^2$ & $0$ & $1$ & $1$ & $0$ & $T_{\epsilon}$ & $0$ & $T_{\epsilon}$ 
			& $0$ & $T_{\epsilon}^{-1}$ &$ 0$ & $0$ & $1$ &$ 0$ \\
			$2_{v}$&	$T_{\epsilon}^2$ & $T_{\epsilon}^2$ & $0$ &$ 1$ & $T_{\epsilon}^3 $& $0$ & $0$ 
			& $T_{\epsilon}$ & $T_{\epsilon}$ & $0$ & $0$ & $T_{\epsilon}^2$ & $0$ &$ 0$ \\
			$3_{v}$&	$T_{\epsilon}^2 $&$ T_{\epsilon}^2$ & $T_{\epsilon}^2$ & $0$ & $T_{\epsilon}^3 $
			&$ T_{\epsilon}^3 $& $T_{\epsilon}^3$ &$ 0$ & $0$ & $0$ & $T_{\epsilon}^4$ & $0$ & $0$ & $0$ \\
			\hline	\hline	
			$0_{u}$&	$0$ & $0$ & $T_{\epsilon}^{-1}$ & $T_{\epsilon}^{-1}$ & $0$ & $1$ & $1$ & $1$ & $1$ 
			& $1+T_{\epsilon}^{-2}$ & $0$ & $0$ & $T_{\epsilon}^{-1}$ & $T_{\epsilon}^{-1}$ \\
			$1_{u}$&	$0$ & $T_{\epsilon}$ & $0$ & $T_{\epsilon}^{-1}$ & $T_{\epsilon}^2$ & $0$ & $1$ & $1$ 
			& $2$ & $1$ & $0$ & $T_{\epsilon}$ & $0$ & $T_{\epsilon}^{-1}$ \\
			$2_{u}$&	$T_{\epsilon}^3$ & $0$ & $0$ & $T_{\epsilon}^{-1}$ & $T_{\epsilon}^2$ & $T_{\epsilon}^2$
			& $0$ & $T_{\epsilon}^2+1$ & $1$ & $1$ & $0$ & $T_{\epsilon}$ & $T_{\epsilon}$ & $0$ \\
			$\tilde{2}_{u}$&	$0$ & $T_{\epsilon}$ & $T_{\epsilon}$ & $0$ & $T_{\epsilon}^2$ & $T_{\epsilon}^2$ 
			& $T_{\epsilon}^2+1$ & $0$ & $1$ & $1$ & $T_{\epsilon}^3$ & $0$ & $0$ & $T_{\epsilon}^{-1}$ \\
			$3_{u}$&	$T_{\epsilon}^3$ & $0$ & $T_{\epsilon}$ & $0$ & $T_{\epsilon}^2$ & $2$ $T_{\epsilon}^2$
			& $T_{\epsilon}^2$ & $T_{\epsilon}^2$ & $0$ & $1$ & $T_{\epsilon}^3$ & $0$ & $T_{\epsilon}$ & $0$ \\
			$4_{u}$&	$T_{\epsilon}^3$ & $T_{\epsilon}^3$ & $0$ & $0$ & $T_{\epsilon}^4+T_{\epsilon}^2$ 
			& $T_{\epsilon}^2$ & $T_{\epsilon}^2$ & $T_{\epsilon}^2$ & $T_{\epsilon}^2$ & $0$ & $T_{\epsilon}^3$ & $T_{\epsilon}^3$
			& $0$ & $0$ \\
			\hline	\hline	
			$0_{\bar{v}}$&	$0$ & $0$ & $0$ & $T_{\epsilon}^{-2}$ & $0$ & $0$ & $0$ & $T_{\epsilon}^{-1}$ & $T_{\epsilon}^{-1}$ &$T_{\epsilon}^{-1}$ & $0 $& $1$ & $1$ & $1$ \\
			$1_{\bar{v}}$&	$0$ & $0$ & $1$ & $0$ & $0$ & $T_{\epsilon}$ & $T_{\epsilon}$ & $0$ & $0$ 
			& $T_{\epsilon}^{-1}$ & $T_{\epsilon}^2$ & $0$ & $1$ & $1$ \\
			$2_{\bar{v}}$&	$0$ & $T_{\epsilon}^2$ & $0$ & $0$ & $T_{\epsilon}^3$ & $0$ & $T_{\epsilon}$ & $0$ 
			& $T_{\epsilon}$ &$ 0$ & $T_{\epsilon}^2$ &$ T_{\epsilon}^2$ & $0$ & $1$ \\
			$3_{\bar{v}}$&	$T_{\epsilon}^4$ & $0$ & $0$ & $0$ & $T_{\epsilon}^3$ & $T_{\epsilon}^3$ & $0$ 
			& $T_{\epsilon}^3$ & $0$ & $0$ & $T_{\epsilon}^2$ & $T_{\epsilon}^2$ & $T_{\epsilon}^2$ & $0$ \\
		\end{tabular}
	}
	\end{center}
\caption{Gives $X(T_a,\lambda|T_b,\mu)$ with $T_aT^{-1}_b$ suppressed for example  $X(T_{v_i},2_v|T_{u_j},0_{u})=T_{v_i}T^{-1}_{u_j}T_{\epsilon}^3$.}
\label{gen_CH_table}
\end{table}
\subsection{Representation of $\Pi^{(N_{\psi}N_{\phi}N_{\bar{\psi}})}_{mat}$ as a sum over  diagrams}
\label{gen_main}
Due to localization for given $N_{\phi}$, $N_{\psi}$ and $N_{\bar{\psi}}$ the matrix part of the hexagonal Wl is
a sum over fixed points: 
\bea \label{PIYoung_gen}
\Pi^{(N_{\psi}N_{\phi}N_{\bar{\psi}})}_{f}=\sum_{\vec{Y}}F_{\vec{Y}}\,.
\eea
Specifically  the  sum is  over all collections of $N_{\phi}+N_{\psi}+N_{\bar{\psi}}$ diagrams of the form 
 $\vec{Y}\equiv\{Y^v_1,Y^v_2,...,Y^v_{N_{\psi}};Y^u_1,Y^u_2,...,Y^u_{N_{\phi}};Y^{\bar{v}}_1,Y^{\bar{v}}_2,...,Y^{\bar{v}}_{N_{\bar{\psi}}}\}$ with entries taken  from the  list given in Fig. \ref{posibleYD_gen}. As the notation indicates $Y^v_i$, $Y^u_i$ and $Y^{\bar{v}}_i$ are diagrams taken from the first second or third row in Fig.\ref{posibleYD_gen} respectively.
In $\vec{Y}$ the total numbers of boxes on the upper, middle and lower lines   are  $K_1$,  $K_2$ and  $K_3$ respectively.
 
The summands  $F_{\vec{Y}}$ are given by 
\begin{small}
	\bea \label{gen_Fwil}
	1/F_{\vec{Y}}=
	\prod_{i,j=1}^{N_{\psi}}\prod_{m,n=1}^{N_{\phi}}\prod_{k,l=1}^{N_{\bar{\psi}}}
	P(v_i,Y^{v}_i|v_j,Y^{v}_j)P(v_i,Y^{v}_i|u_n,Y^{u}_n)P(v_i,Y^{v}_i|\bar{v}_k,Y^{\bar{v}}_k)
	\\ \nonumber
	\times P(u_n,Y^u_n|v_i,Y^{v}_i) P(u_n,Y^u_n|u_m,Y^u_m)	 P(u_n,Y^u_n|\bar{v}_k,Y^{\bar{v}}_k)
	 \\ \nonumber 
	 \times	P(\bar{v}_k,Y^{\bar{v}}_k|v_i,Y^{v}_i)P(\bar{v}_k,Y^{\bar{v}}_k|u_n,Y^u_n)
	P(\bar{v}_k,Y^{\bar{v}}_k|\bar{v}_l,Y^{\bar{v}}_l)
	\eea
\end{small}
where $P(a,\lambda|b,\mu)$ are displayed  in Table \ref{gen_table:2}.
\begin{table} [h!]
	\begin{center}
		\resizebox{18cm}{!}{
			\begin{tabular}{| l || c | c | c |c|| c | c | c | c | c | c || c| c |c| c | r| }
				\hline
&$0_v$ & $1_v$ & $2_v$ &$3_v$	&$0_u$ & $1_u$ & $2_u$ &$\tilde{2}_u$&$3_u$&$4_u$&$0_{\bar{v}}$&$1_{\bar{v}}$ &$2_{\bar{v}}$&$3_{\bar{v}}$ \\ \hline\hline
$0_v$  &$1$ & $v_{ij}$ & $v_{ij}$ &$v_{ij}$
&$1$&$1$&$v_i-u_j-\frac{i}{2}$&$1$&$v_i-u_j-\frac{i}{2}$&$v_i-u_j-\frac{i}{2}$
&$1$&$1$&$1$&$v_i-\bar{v}_j-i$ \\ \hline
$1_v$  &$v_{ij}+i$ &$ 1$ & $v_{ij}$ &$v_{ij}$
&$1$&$v_i-u_j+\frac{i}{2}$&$1$&$v_i-u_j+\frac{i}{2}$&$1$&$v_i-u_j-\frac{i}{2}$
&$1$&$1$ &$v_i-\bar{v}_j$&$1$\\ \hline
$2_v$  &$v_{ij}+i$ & $v_{ij}+i$ & $1$ &$v_{ij}$
&$v_i-u_j+\frac{3i}{2}$&$1$&$1$&$v_i-u_j+\frac{i}{2}$&$v_i-u_j+\frac{i}{2}$&1
&$1$&$v_i-\bar{v}_j+i$ &$1$&$1$\\ \hline
$3_v$  &$v_{ij}+i$ & $v_{ij}+i$ & $v_{ij}+i$ &$1$
&$v_i-u_j+\frac{3i}{2}$&$v_i-u_j+\frac{3i}{2}$&$v_i-u_j+\frac{3i}{2}$&$1$&$1$&$1$
&$v_i-\bar{v}_j+2i$&$1$&$1$&$1$ \\ \hline\hline
			$0_u$&$1$&$1$&$u_i-v_j-\frac{i}{2}$&$u_i-v_j-\frac{i}{2}$
				&$1$ & $u_{ij}$ & $u_{ij}$ &$u_{ij}$&$u_{ij}$&$u_{ij}(u_{ij}-i)$ 
			&$1$&$1$&$u_i-\bar{v}_j-\frac{i}{2}$&$u_i-\bar{v}_j-\frac{i}{2}$	
			\\ \hline
				$1_u$	&$1$&$u_i-v_j+\frac{i}{2}$&$1$&$u_i-v_j-\frac{i}{2}$
				&$u_{ij}+i$ & $1$ & $u_{ij}$ &$u_{ij}$&$u_{ij}^2$&$u_{ij}$
					&$1$&$u_i-\bar{v}_j+\frac{i}{2}$&$1$&$u_i-\bar{v}_j-\frac{i}{2}$
				 \\ \hline
				$2_u$ &$u_i-v_j+\frac{3i}{2}$&$1$&$1$&$u_i-v_j-\frac{i}{2}$
					&$u_{ij}+i$ & $u_{ij}+i$ & $1$ &$u_{ij}(u_{ij}+i)$&$u_{ij}$&$u_{ij}$
				&$1$&$u_i-\bar{v}_j+\frac{i}{2}$&$u_i-\bar{v}_j+\frac{i}{2}$&$1$
					 \\ \hline
				$\tilde{2}_u$ &$1$&$u_i-v_j+\frac{i}{2}$&$u_i-v_j+\frac{i}{2}$&$1$
					&$u_{ij}+i$ & $u_{ij}+i$ & $u_{ij}(u_{ij}+i)$ &$1$&$u_{ij}$&$u_{ij}$
				&$u_i-\bar{v}_j+\frac{3i}{2}$&$1$&$1$&$u_i-\bar{v}_j-\frac{i}{2}$	
					 \\ \hline
				$3_u$ &$u_i-v_j+\frac{3i}{2}$&$1$&$u_i-v_j+\frac{i}{2}$&$1$
					&$u_{ij}+i$ & $(u_{ij}+i)^2$ & $u_{ij}+i$ &$u_{ij}+i$&$1$&$u_{ij}$ 
					&$u_i-\bar{v}_j+\frac{3i}{2}$&$1$&$u_i-\bar{v}_j+\frac{i}{2}$&$1$
					\\ \hline
				$4_u$&$u_i-v_j+\frac{3i}{2}$&$u_i-v_j+\frac{3i}{2}$&$1$&$1$
					&$(u_{ij}+i)(u_{ij}+2i)$ & $u_{ij}+i$ & $u_{ij}+i$ &$u_{ij}+i$&$u_{ij}+i$&$1$
				&$u_i-\bar{v}_j+\frac{3i}{2}$&$u_i-\bar{v}_j+\frac{3i}{2}$&$1$&$1$	
					 \\ \hline \hline
$0_{\bar{v}}$	&$1$ & $1$ & $1$ &$\bar{v}_i-v_j-i$
&$1$&$1$&$1$&$\bar{v}_i-u_j-\frac{i}{2}$&$\bar{v}_i-u_j-\frac{i}{2}$&$\bar{v}_i-u_j-\frac{i}{2}$
&$1$&$\bar{v}_{ij}$&$\bar{v}_{ij}$&$\bar{v}_{ij}$ \\ \hline
$1_{\bar{v}}$	&$1$ & $1$  &$\bar{v}_i-v_j$&$1$
&$1$&$\bar{v}_i-u_j+\frac{i}{2}$&$\bar{v}_i-u_j+\frac{i}{2}$&$1$&$1$&$\bar{v}_i-u_j-\frac{i}{2}$
&$\bar{v}_{ij}+i$&$1$&$\bar{v}_{ij}$&$\bar{v}_{ij}$ \\ \hline
$2_{\bar{v}}$	&$1$ & $\bar{v}_i-v_j+i$ & $1$ &$1$
&$\bar{v}_i-u_j+\frac{3i}{2}$&$1$&$\bar{v}_i-u_j+\frac{i}{2}$&$1$&$\bar{v}_i-u_j+\frac{i}{2}$&$1$
&$\bar{v}_{ij}+i$&$\bar{v}_{ij}+i$&$1$&$\bar{v}_{ij}$ \\ \hline
$3_{\bar{v}}$	&$\bar{v}_i-v_j+2i$ & $1$ & $1$ &$1$
&$\bar{v}_i-u_j+\frac{3i}{2}$&$\bar{v}_i-u_j+\frac{3i}{2}$&$1$&$\bar{v}_i-u_j+\frac{3i}{2}$&$1$&$1$
&$\bar{v}_{ij}+i$&$\bar{v}_{ij}+i$&$\bar{v}_{ij}+i$&$1$ \\ \hline				
			\end{tabular}
		}
	\end{center}
	\caption{This table gives  $P(v_i,Y^{v}_i|v_j,Y^{v}_j)$, $P(v_i,Y^{v}_i|\bar{v}_j,Y^{\bar{v}}_j)$,  $P(v_i,Y^{v}_i|u_j,Y^{u}_j)$, 
		$P(\bar{v}_i,Y^{\bar{v}}_i|v_j,Y^{v}_j)$, $P(\bar{v}_i,Y^{\bar{v}}_i|\bar{v}_j,Y^{\bar{v}}_j)$... for all diagrams that is $Y^{v}_i\in \{0_v,1_v,2_v,3_v\}$, $Y^{\bar{v}}_i\in \{0_{\bar{v}},1_{\bar{v}},2_{\bar{v}},3_{\bar{v}}\}$ and $Y^{u}_i\in \{0_u,1_u,2_u,\tilde{2}_u,3_u,4_u\}$}
	\label{gen_table:2}
\end{table}
\section{Explicit results for $\Pi^{(N_{\psi}N_{\phi}N_{\bar{\psi}})}_{mat}$}
\label{GEN_resulats}
Examining numerous cases (some of them are presented in subsections \ref{rb=1}-\ref{rb=4})
we got convinced that the denominator of 
\bea
\Pi^{(N_{\psi}N_{\phi}N_{\bar{\psi}})}_{mat}=\frac{N^{(N_{\psi}N_{\phi}N_{\bar{\psi}})}}
{D^{(N_{\psi}N_{\phi}N_{\bar{\psi}})}}
\eea 
has the form
\bea
\label{den}
D^{(N_{\psi}N_{\phi}N_{\bar{\psi}})}=
\prod_{i=1}^{N_{\psi}}\prod_{k=1}^{N_{\phi}}
\left(4(u_k-v_i)^2+9\right)
\prod_{k=1}^{N_{\phi}}\prod_{m=1}^{N_{\bar{\psi}}}
\left(4(u_k-\bar{v}_m)^2+9\right)\times\\\nonumber
\prod_{i=1}^{N_{\psi}}\prod_{m=1}^{N_{\bar{\psi}}}
\left((v_i-\bar{v}_m)^2+4\right)
\prod_{i>j}^{N_{\psi}}\left(v^2_{ij}+1\right)
\prod_{m>n}^{N_{\bar{\psi}}}\left(\bar{v}^2_{mn}+1\right)
\prod_{k>l}^{N_{\phi}}\left(u_{kl}^2+1\right)\left(u_{kl}^2+4\right)\,.
\eea
We have succeeded  to  generalize  the  recursion properties (\ref{rec}) and (\ref{recF}):
\bea 
\label{rec_gen_sc}
\nonumber
\textit{Res}_{u_2=u_1+2i}\Pi_{mat}^{(N_{\psi}N_{\phi}N_{\bar{\psi}})}(u_1,\cdots ,u_{N_{\phi}}) =
(-)^{ m_1+m_2+1}\bigg(2i\displaystyle\prod_{j=3}^{N_{\phi}}u_{1j}(u_{1j}+i)^2(u_{1j}+2i)\times \\ 
	\prod^{N_{\psi}}_{m=1}(v_m-u_1-\frac{i}{2})(u_1-v_m+\frac{3i}{2})
\prod^{N_{\bar{\psi}}}_{n=1}(\bar{v}_n-u_1-\frac{i}{2})(u_1-\bar{v}_n+\frac{3i}{2})\bigg)^{-1} 
\times\quad
\\ \nonumber
\Pi_{mat}^{(N_{\psi},N_{\phi}-2,N_{\bar{\psi}})}(u_3,\cdots ,u_{N_{\phi}})
\eea
 and 
 	\bea
 	\label{recF_gen}
 	&i\textit{Res}_{\bar{v}_1=v_1+2i}\Pi_{mat}^{(N_{\psi}N_{\phi}N_{\bar{\psi}})}(v_1,\cdots ,v_{N_{\psi}};\bar{v}_1,...,\bar{v}_{N_{\bar{\psi}}}) = \\ \nonumber
 	& \frac{(-)^{m_1+m_2+1} \Pi_{mat}^{(N_{\psi}-1,N_{\phi},N_{\bar{\psi}}-1)}
 	 	(v_2,\cdots ,v_{N_{\psi}};\bar{v}_2,...,\bar{v}_{N_{\bar{\psi}}})}
 	{\displaystyle\prod_{k=2}^{N_{\psi}}(v_{1k}+i)v_{1k}
 		\displaystyle\prod_{j=2}^{N_{\bar{\psi}}}(v_1-\bar{v}_j+2i)(v_1-\bar{v}_j+i)
 	\prod_{n=1}^{N_{\phi}}\big(v_1-u_n+\frac{3i}{2}\big)\big(u_n-v_1-\frac{i}{2}\big)}\,,
 	\eea
 	where $ m_{1,2}$ are   defined  in (\ref{m}) for pairs $\Pi_{mat}$ related by above
 	recursions.
 	We do not present  proofs of these recursion relations, since structurally they are  similar to the proofs of simpler cases considered in sections \ref{scalars} and \ref{fermion}.
 	
 	In the upcoming subsections we will use our combinatorial formula to compute $\Pi_{mat}^{(N_{\psi}N_{\phi}N_{\bar{\psi}})}$ explicitly for some fixed values of $r_b$, $N_{\psi}$,
 	 $N_{\phi}$, $N_{\bar{\psi}}$. 
 	 For some cases\footnote{Those with $K_1+K_2+K_3\le 4$.} we have checked our results against 
 	 (\ref{gen_Pmat1}) by performing  numerical integrations.

\subsection{Cases with $r_b=1$}
\label{rb=1}
\begin{itemize}
	\item If  $N_{\psi }=3$, $N_{\phi }=0$ and  $N_{\bar{\psi }}=0$, according to (\ref{res1})-(\ref{res3}) we have
	 $K_1 =2$, $K_2=1$ and $K_3=0$. There are $6$
	 fixed points $\{0_v,1_v,2_v\}$, $\{0_v,2_v,1_v\}$, $\{1_v,0_v,2_v\}$, $\{1_v,2_v,0_v\}$, $\{2_v,0_v,1_v\}$ and $\{2_v,1_v,0_v\}$, thus using our formula (\ref{PIYoung_gen}) and (\ref{piF_piMAT}) we get
	\bea
	\Pi_{mat}^{(300)}=\frac{6}{\left(v_{12}^2+1\right) \left(v_{13}^2+1\right) \left(v_{23}^2+1\right)}\,.
	\eea
	\item When $N_{\psi }=0$ ,  $N_{\phi }=2$ and $N_{\bar{\psi }}=1$ we have $K_1=1$ $K_2=2$ and $K_3=1$.
	Here there are $12$ fixed points $\{0_u,1_u,3_{\bar{v}}\}$, $\{0_u,2_u,2_{\bar{v}}\}$, $\{0_u,4_u,0_{\bar{v}}\}$,
	 $\{1_u,0_u,3_{\bar{v}}\}$, 	$\{1_u,2_u,1_{\bar{v}}\}$,
	$\{1_u,3_u,0_{\bar{v}}\}$,	$\{2_u,0_u,2_{\bar{v}}\}$, $\{2_u,1_u,1_{\bar{v}}\}$, $\{2_u,\tilde{2}_u,0_{\bar{v}}\}$, 
	$\{\tilde{2}_u,2_u,0_{\bar{v}}\}$, $\{3_u,1_u,0_{\bar{v}}\}$, $\{4_u,0_u,0_{\bar{v}}\}$
	\bea
	\Pi_{mat}^{(021)}=\frac{24\left(-4 e_1 \bar{v}+6 e_1^2-20 e_2+4 \bar{v}^2+45\right)}
	{\left(u_{12}^2+1\right) \left(u_{12}^2+4\right) \left(4 \left(\bar{v}-u_1\right){}^2+9\right) \left(4 \left(\bar{v}-u_2\right){}^2+9\right)}\,,
	\eea
	where $e_1$ and $e_2$  are the elementary symmetric polynomials in $u_1$ and $u_2$.
	\item The case with $N_{\psi }=2$, $N_{\phi }=1$ and $N_{\bar{\psi }}=1$ thus $K_1=K_2=2$ and  $K_3=1$. We have $27$
	fixed points and the final result is
	\bea
	N^{(211)}=96\left(-48 s_1 u+12 u^2 \bar{v}^2+80 u^2-8 u \bar{v}^3-64 u\bar{v}+8 \bar{v}^4+143 \bar{v}^2+612\right.\\
	\nonumber
	-12 s_1 u^2 \bar{v}+8 s_1^2 u^2-8 s_1^2 u \bar{v}-12 s_1\bar{v}^3+12 s_1^2 \bar{v}^2-111 s_1 \bar{v}+90 s_1^2\quad\\
	\nonumber
	\left.-20 s_2 u^2+56 s_2 u \bar{v}-8 s_1 s_2 u-12 s_2 \bar{v}^2-16 s_1 s_2 \bar{v}+12 s_2^2-201 s_2\right),
	\eea
	where $s_1$ and $s_2$  are the elementary symmetric polynomials in $v_1$ and $v_2$.
	 The  denominator is given by (\ref{den}).
	\item The case with $N_{\psi }=4$, $N_{\phi }=0$ and  $N_{\bar{\psi }}=1$ so  $K_1=3$, $K_2=2$ and $K_3=1$ this case
	has $60$ fixed points. Our formula gives
	\bea
	N^{(401)}=12\left(-3 s_1 \bar{v}^5-49 s_1 \bar{v}^3-196 s_1 \bar{v}\nonumber
	+2 \bar{v}^6+49 \bar{v}^4+392 \bar{v}^2+1040\right.\\\nonumber
	+3 s_1^2 \bar{v}^4-3 s_2 \bar{v}^4+42 s_1^2 \bar{v}^2
	-63 s_2 \bar{v}^2-28 s_1 s_2\bar{v}+152 s_1^2-340 s_2\\\nonumber
	-4 s_1 s_2 \bar{v}^3+14 s_3 \bar{v}^3+3 s_2^2 \bar{v}^2
	-3 s_1 s_3 \bar{v}^2+119 s_3 \bar{v}+20 s_2^2-46 s_1 s_3\\
	\left.-30 s_4 \bar{v}^2-3 s_2 s_3 \bar{v}+15 s_1 s_4 \bar{v}+2 s_3^2-5 s_2 s_4+65 s_4\right).\qquad
	\eea
Here the $s_i$ are  elementary symmetric polynomials $v_i$, $i=1,...,4$. For the  denominator see (\ref{den}).
	\item The case with $N_{\psi }=0$, $N_{\phi }=0$ and $N_{\bar{\psi }}=5$ thus $K_1=1$, $K_2=2$ and $K_3=3$.
	 We have $60$ fixed points and the final result is 
	\bea
	\Pi_{mat}^{(005)}=\frac{12 \left(2 \bar{s}_1^2-5 \bar{s}_2+25\right)}
	{ \prod _{i>j}^5 \left(\bar{v}_{ij}^2+1\right)}\,,
	\eea
	where $\bar{s}_i$ are the elementary symmetric polynomials $\bar{v}_i$, $i=1,...,5$.
\end{itemize}
\subsection{Cases with $r_b=2$}
 \begin{itemize}
 		\item 	The case $ N_{\psi }=2$, $N_{\phi }=0$ and $ N_{\bar{\psi }}=0 $ thus $K_1=1$,  $K_2=K_3=0$.
 		 We have only 2 fixed points $\{0_v,1_v\}$ and $\{1_v,0_v\}$ using our formula we get
 	\bea
 	\Pi_{mat}^{(200)}=\frac{2}{v_{12}^2+1}\,.
 	\eea
 	\item The case  $N_{\phi }=3$, $N_{\psi }=N_{\bar{\psi }}=0$ so  $K_1=K_3=1$ and $K_2=2$.
 	 Here we have $15$ fixed points which are: $\{0_u,0_u,4_u\}$, $\{0_u,1_u,3_u\}$,
 	$\{0_u,2_u,\tilde{2}_u\}$, $\{0_u,\tilde{2}_u,2_u\}$, $\{0_u,3_u,1_u\}$, $\{0_u,4_u,0_u\}$, $\{1_u,0_u,3_u\}$,
 	$\{1_u,3_u,0_u\}$, $\{2_u,0_u,\tilde{2}_u\}$, $\{2_u,\tilde{2}_u,0_u\}$, $\{\tilde{2}_u,0_u,2_u\}$,
 	$\{\tilde{2}_u,2_u,0_u\}$, $\{3_u,0_u,1_u\}$, $\{3_u,1_u,0_u\}$, $\{4_u,0_u,0_u\}$. Using
 	(\ref{PIYoung_gen}), (\ref{piF_piMAT}), (\ref{gen_Fwil}) and Table \ref{gen_table:2} we get
 	\bea
 	\Pi^{(030)}=\frac{6 \left(e_1^2-3 e_2+7\right) \left(e_1^2-3 e_2+12\right)}{\prod_{i<j}(u_{ij}^2+1)(u_{ij}^2+4)}\,,
 	\eea
 	where  $e_1$ and $e_2$  are the elementary symmetric polynomials in $u_1$, $u_2$ and $u_3$.
 	This coincides with  \cite{Basso:2015uxa}.
 		\item Case  $N_{\psi }=N_{\phi }=N_{\bar{\psi }}=1$ so  $K_1=K_2=K_3=1$ has $8$ fixed points:
 	$\{0_v,  0_u,3_{\bar{v}}\}$, $\{0_v, 2_u, 1_{\bar{v}}\}$, $\{0_v, 3_u, 0_{\bar{v}}\}$, $\{1_v,  0_u, 2_{\bar{v}}\}$, $\{1_v, 1_u,1_{\bar{v}}\}$, $\{1_v,\tilde{2}_u,0_{\bar{v}}\}$,
 	$\{2_v,0_u, 1_{\bar{v}}\}$, $\{3_v, 0_u, 0_{\bar{v}}\}$. 
 	Using our formula we get
 	\bea
 	\Pi_{mat}^{(111)}=\frac{16 \left(4 u^2-4 u v-4 u \bar{v}+6 v^2-8 v \bar{v}+6 \bar{v}^2+45\right)}
 	{\left(4 (u-v)^2+9\right) \left(4 (u-\bar{v})^2+9\right) \left((v-\bar{v})^2+4\right)}\,.
 	\eea
 		\item Case $ N_{\psi }=N_{\phi }=2,N_{\bar{\psi }}=0 $ thus  $K_1=K_2=2,K_3=1$ has $34$ fixed points,  we get
 	\begin{footnotesize}
 		\bea
 		&&N^{(220)}=192\left(-16 e_1^3 s_1-148 e_1 s_1+16 e_1^4-96 e_2 e_1^2
 		+316 e_1^2+144 e_2^2-968 e_2+1449\right.\,\\
 		\nonumber
 		&&\left.+24 e_1^2 s_1^2-48 e_1^2 s_2+48 e_2 e_1 s_1-16 e_1 s_1 s_2-80 e_2 s_1^2
 		+224 e_2 s_2+180 s_1^2+16 s_2^2-424 s_2\right).
 		\eea 
 	\end{footnotesize}
 	Here $e_1$ and $e_2$ ($s_1$ and $s_2$) are the elementary symmetric polynomials in $u_1$ and $u_2$  ($v_1$ and $v_2$).
 	The denominator is given by (\ref{den}).
 		\item	Case $ N_{\psi }=6$, $N_{\phi }=0$ and $N_{\bar{\psi }}=0$ so $K_1=4$, $K_2=2$ $K_3=1$
 		 has $180$ fixed points, we get
 	\bea
 	\Pi^{(600)}_{mat}=\frac{24 \left(25 s_1^2-5 s_3 s_1+2 s_2^2-60 s_2+10 s_4+210\right)}
 	{ \prod _{i>j}^6 \left(v_{ij}^2+1\right)}\,,
 	\eea
 	where $s_i$ are elementary symmetric polynomials in $v_i$, $i=1,...,6$.
 \end{itemize}
\subsection{Cases with $r_b=3$}
\begin{itemize}
	\item The case with $N_{\psi }=N_{\phi }=0$ and  $N_{\bar{\psi }}=3$ thus
	$K_1=0$ $K_2=1$ and  $K_3=2$. Here we have $6$ fixed points
	$\{0_{\bar{v}},1_{\bar{v}},2_{\bar{v}}\}$, $\{0_{\bar{v}},2_{\bar{v}},1_{\bar{v}}\}$, 
	$\{1_{\bar{v}},0_{\bar{v}},2_{\bar{v}}\}$, $\{1_{\bar{v}},2_{\bar{v}},0_{\bar{v}}\}$,
	 $\{2_{\bar{v}},0_{\bar{v}},1_{\bar{v}}\}$, $\{2_{\bar{v}},1_{\bar{v}},0_{\bar{v}}\}$,
	 using 	 (\ref{PIYoung_gen}), (\ref{gen_Fwil}) and Table \ref{gen_table:2} we get
	 \bea
	 \Pi_{mat}^{(003)}=
	 \frac{6}{\left(\bar{v}_{12}^2+1\right) \left(\bar{v}_{13}^2+1\right) \left(\bar{v}_{23}^2+1\right)}\,.
	 \eea
	 \item Case with $N_{\psi }=1$, $N_{\phi }=2$ and  $N_{\bar{\psi }}=0$ thus
	 $k_1=1$ $k_2=2$  and $k_3=1$ has $12$ fixed points. Our result is
	 \bea
	 \Pi_{mat}^{(120)}=\frac{24 \left(-4 e_1 v+6 e_1^2-20 e_2+4 v^2+45\right)}
	 {\left(u_{12}^2+1\right) \left(u_{12}^2+4\right) 
	 	\left(4 \left(u_1-v\right){}^2+9\right) \left(4 \left(u_2-v\right){}^2+9\right)}\,,
	 \eea
	 where $e_1$ and $e_2$  are the elementary symmetric polynomials in $u_1$, $u_2$.
	 	 \item Case with $N_{\psi }=1$, $N_{\phi }=1$ and  $N_{\bar{\psi }}=2$ therefor $K_1=1$ and  $K_2=K_3=2$ has
	 	 $27$ fixed points. We get 
	 	 \begin{small}
	 	 	\bea
	 	 	\nonumber
	 	 	N^{(112)}=-96\left(12 \bar{s}_1 u^2 v-8 \bar{s}_1^2 u^2+20 \bar{s}_2 u^2+48 \bar{s}_1 u-90 \bar{s}_1^2+201 \bar{s}_2-80 u^2-612\right.  \\ \nonumber
	 	 	+8 \bar{s}_1^2 u v-56 \bar{s}_2 u v+
	 	 	8\bar{s}_1 \bar{s}_2 u+111 \bar{s}_1 v+16 \bar{s}_1 \bar{s}_2 v-12 \bar{s}_2^2+64 u v\\
	 	 	+12 \bar{s}_1 v^3-12 \bar{s}_1^2 v^2+12 \bar{s}_2 v^2
	 	 	-12 u^2 v^2+8 u v^3-8 v^4-143 v^2,\quad
	 	 	\eea
	 	 \end{small}
 	  $\bar{s}_1$ and $\bar{s}_2$  are the elementary symmetric polynomials in $\bar{v}_1$, $\bar{v}_2$.
 	  For the  denominator see (\ref{den}).
 	  \item Case with $N_{\psi}=5$, $N_{\phi}=0$ and $N_{\bar{\psi}}=0$ so $K_1=3$, $K_2=2$ and $K_3=1$
 	  has $60$ fixed points. The result is
 	  \bea
 	 \Pi_{mat}^{(500)} =\frac{24 s_1^2-60 s_2+300}{\prod_{i>j}^5(v_{ij}^2+1)}\,,
 	  \eea
 	  where $s_i$ are elementary symmetric polynomials in $v_i$, $i=1,...,5$.
\end{itemize}

\subsection{Cases with $r_b=0$ or $r_b=4$}
\label{rb=4} 
	As one can see the constraints (\ref{res1})-(\ref{res3}) on $K_1$, $K_2$ and $K_3$ are the same when  $r_b=0$ and $r_b=4$
	so this two cases give rise to the same $\Pi^{(N_{\psi}N_{\phi}N_{\bar{\psi}})}_{mat}$. 
\begin{itemize}
	\item Case with $N_{\psi }=2$, $N_{\phi }=1$ and  $N_{\bar{\psi }}=0$ so $k_1=2$, $k_2=2$ and $k_3=1$ has $12$
	fixed points. The result is
	\bea
	\Pi^{(210)}_{mat}=\frac{192}{\left(v_{12}^2+1\right) 
		\left(4 \left(u-v_1\right){}^2+9\right) \left(4 \left(u-v_2\right){}^2+9\right)}\,.
	\eea
	\item Case with $N_{\psi }=1$, $N_{\phi }=2$ and  $N_{\bar{\psi }}=1$ so
	$K_1=2$,  $K_2=3$ and  $K_3=2$ has $48$ fixed points. The result is
	\bea
	N^{(121)}=384(-84 e_1 v-84 e_1 \bar{v}+16 v^2 \bar{v}^2
	+180 v^2-192 v \bar{v}+180 \bar{v}^2+1161\quad\\ \nonumber
	-16 e_1 v^2 \bar{v}+24 e_1^2 v^2-16 e_1 v \bar{v}^2-32 e_1^2 v \bar{v}
	+24 e_1^2 \bar{v}^2+180 e_1^2-80 e_2 v^2\quad \\ \nonumber
	+192 e_2 v \bar{v}-16 e_1 e_2 v-80 e_2 \bar{v}^2-16 e_1 e_2 \bar{v}+16 e_2^2-552 e_2),\quad
	\eea
	where $e_1$ and $e_2$ are the elementary symmetric polynomials $u_1$ and $u_2$. 
	For the  denominator see (\ref{den}).
	\item Case with  $N_{\psi }=4$, $N_{\phi }=0$ and $N_{\bar{\psi }}=0$  thus 
	$K_1=3$, $K_2=2$ and $K_3=1$ has $24$ fixed points. The result is
	\bea
	\Pi_{mat}^{(400)}=\frac{24}{\prod_{i>j}^{4}(v_{ij}^2+1)}\,.
	\eea
\end{itemize}
\section{Conclusions and perspectives}
It would be interesting to find a physical interpretation of the ADHM-like moduli space we have constructed. Perhaps the identification of  the real ADHM equation counterpart of the matrix equations (\ref{ADHMa}-\ref{ADHMd}) would have some significance as well. Another achievement of this paper may be consider that of a general approach for passing from an integral representation with some group-theoretical structure to a combinatorial sum over Young diagrams.

Thanks to the peculiar two $SU(4)$ matrix structure of the fermions and anti-fermions contributions, the papers \cite{FPR1,BFPR4, BFPR1} have re-summed the leading contributions to those from gluons (and their bound states) at strong coupling so that to give the same Thermodynamic Bethe Ansatz results as (classical) string theory (\cite{BFPR4} furnishes by the same method also  subleading corrections, waiting for one-loop confirmation). This computation resembles the poles contributions of the instanton partition function of $\mathcal{N}=2$ SYM \cite{Nekrasov:2002qd} in the so-called Nekrasov-Shatashvili (NS) limit \cite{NekSha} (and similarly for the subleading correction to the NS limit, computed in \cite{BouFio1,BouFio2}). If this represents a second way to TBA (with respect to the ordinary one \cite{Zam-TBA})), which surprisingly stems from FFs, a third one can be counted as the massive Ordinary Differential Equation/ Integrable Model (ODE/IM) correspondence which \cite{FRShu} has recently shown to 'solve' the dual (classical) string theory thanks to the full-fledged quantum integrability structures: not only $T$-, $Y$-systems and TBA \cite{TBuA, YSA, Hatsuda:2010cc}, but also the more fundamental $Q$-functions with relative functional and integral equations. These structures have been derived from the discrete ($\Omega$- and $\Lambda$-) symmetries acting on the gauge (or differential equation) moduli space (with the extension of a twist or angular momentum for the string solution). In this perspective, the FF series is incorporated in a full integrability structure in its strong coupling and hence the (exact) all coupling expressions, we are dealing with here, will acquire even more importance as a possible (second) quantization of the massive ODE/IM correspondence. In a nutshell, the extension (at all couplings) and r\^ole of the $Q$-functions shall be punctually scrutinized as these are the most fundamental objects on the integrability side and the closest to the ODE wave function according to \cite{Fioravanti:2021bzq}. 

Yet, the OPE series is much more complicated than the NS one and, in particular, its strong coupling seems insensible to many details of the weak or all coupling regime (the presence of the scalars, for instance). Maybe this is a positive point in favor of the quantization of massive ODE/IM correspondence as could be argued by looking at the simple and elegant structure of the next to leading expression in the NS regime \cite{BouFio1,BouFio2}. In fact, how to quantize the TBA is a long-standing question, but gauge or string theory may know the answer.

\section*{Acknowledgments}
We acknowledge discussions with S. Penati. This work has been partially supported by the grants: GAST (INFN), the EC Network Gatis and the MIUR-PRIN contract 2017CC72MK\textunderscore 003.


\bibliographystyle{JHEP}
\providecommand{\href}[2]{#2}
\begingroup\raggedright

\endgroup

\end{document}